\documentstyle[referee]{pljour1}
\journalname{The Astronomy and Astrophysics Review}  
\def\cm3{\,{\rm cm^{-3}}} 
\def\cm2{\,{\rm cm^{-2}}}
\def\kms{\,{\rm km\,s^{-1}}} 

\def\Msun{\,{\rm M_\odot}} 
\def\Lsun{\,{\rm L_\odot}} 

\def\h2{\,{\rm H_{2}}}
\def\nH2{\,{\it N(H2)}}
\def\nh{\,{\it N_{\rm H}}}

\def\twcm{\,{\rm 10^{20}\,cm^{-2}}}
 
\def\sun{\odot}

\def\aua{{\rm A\&A} } 
\def\auas{{\rm A\&AS} } 
\def\apj{{\rm ApJ} } 
\def\aj{{\rm AJ} } 
\def\apjs{{\rm ApJS} } 
\def\apjl{{\rm ApJL} } 
\def\araa{{\rm ARAA} } 
\def\mnras{{\rm MNRAS} } 
\def\pasp{{\rm PASP} } 
\def\pasj{{\rm PASJ} }

\begin{document} 
 

\title{Centaurus A -- NGC 5128} 
 
 
\author{F.P. Israel
        \inst{} 
  } 

\institute{Sterrewacht Leiden, P.O. Box 9513, 2300 RA Leiden, the Netherlands 
}  
 
\date{Received: 30 December 1997} 
 
\maketitle 

\begin{abstract} 
At a distance of 3.4 Mpc, NGC 5128 (Centaurus A) is by far the nearest active
radio galaxy. It is often considered to be the prototype Fanaroff-Riley
Class I `low-luminosity' radio galaxy, and as such it plays an important 
role in our understanding of a major class of active galaxies. Its proximity 
has spawned numerous detailed investigations of its properties, yielding 
unrivalled but still incomplete knowledge of its structure and dynamics.

The massive elliptical host galaxy is moderately triaxial and contains a thin, 
strongly warped disk rich in dust, atomic and molecular gas and luminous
young stars. Its globular cluster ensemble has a bimodal distribution of 
metallicities. Deep optical images reveal faint major axis extensions as 
well as a system of filaments and shells. These and other characteristics 
are generally regarded as strong  evidence that NGC 5128 has experienced 
a major merging events at least once in its past.

The galaxy has a very compact, subparsec nucleus exhibiting noticeable
intensity variations at radio and X-ray wavelengths, probably powered by 
accretion events. The central object may be a black hole of moderate mass.
Towards the nucleus, rich absorption spectra of atomic hydrogen and 
various molecular species suggest the presence of significant amounts of 
material falling into the nucleus, presumably `feeding the monster'.
Emanating from the nucleus are linear radio/X-ray jets, becoming
subrelativistic at a few parsec from the nucleus. At about 5 kpc from the
nucleus, the jets expand into plumes. Huge radio lobes extend beyond
the plumes out to to 250 kpc. A compact circumnuclear disk with a central
cavity surrounds the nucleus. Its plane, although at an angle to the minor 
axis of the galaxy, is perpendicular to the inner jets. The jet-collimating
mechanism, probably connected to the circumnuclear disk, appears to 
precess on timescales of order a few times 10$^{7}$ years. 

This review summarizes the present state of knowledge of NGC 5128
and its associated radio source Centaurus A. Underlying physical processes
are outside its scope: they are briefly referred to, but not discussed.


\end{abstract} 

\section{Introduction} 

\subsection{Historical note}

Observing from the Cape of Good Hope, Herschel (1847) was the first to note
the peculiar nature of NGC 5128: {\it `A nebula consisting of two lateral
portions and no doubt of a small streak of nebula along the middle of the
slit or interval between them, having a star at its extremity'} and {\it 
` ... a very problematical object, and must be regarded as forming a genus 
apart, since it evidently differs from mere `double nebulae' ...'}.
Although Hubble (1922) included the galaxy in a list of (mostly galactic) 
nebulae with line emission, it was mostly ignored until the advent of radio 
astronomy about 50 years ago (but see Paraskevoulos 1935). 
Shortly after World War II, Bolton (1948) 
announced the discovery of six discrete sources of 100 MHz radio emission in 
addition to the already known case of Cygnus A. Precisely a year later, using 
their remarkable `cliff-interferometry' technique, Bolton, Stanley $\&$ Slee 
(1949) identified  the sources Taurus A, Virgo A and Centaurus A with the 
nebulous objects M1 (Crab nebula), NGC 4486 (M 87) and NGC 5128 respectively. 
Although they were actually convinced of the extragalactic nature of the 
latter two, their published discussion was noncommittal, reflecting 
referee objections to this notion (R.D. Ekers, private communication). More 
precise measurements by Mills (1952) confirmed the identification of  
Centaurus A with NGC 5128. Discussing the nature of optical 
counterparts to the newly discovered radio sources, Baade $\&$ Minkowski 
(1954) established NGC 5128 as an unusual extragalactic system, located in 
space well `beyond M 31'. They described it as {\it `... an 
unresolved E0 nebula with an unusually strong and wide central dark band, 
a combination highly anomalous for a spherical nebula'} and interpreted it 
as the interaction or merger of an elliptical and a spiral galaxy. 
Nevertheless, the merger hypothesis then receded into the background until 
it was revived some 25 years later by the work of Graham (1979), Dufour et al. 
(1979), Tubbs (1980) and Malin, Quinn $\&$ Graham (1983). It is now the 
favoured explanation for the unusual characteristics of NGC 5128.

\begin{figure}[t]
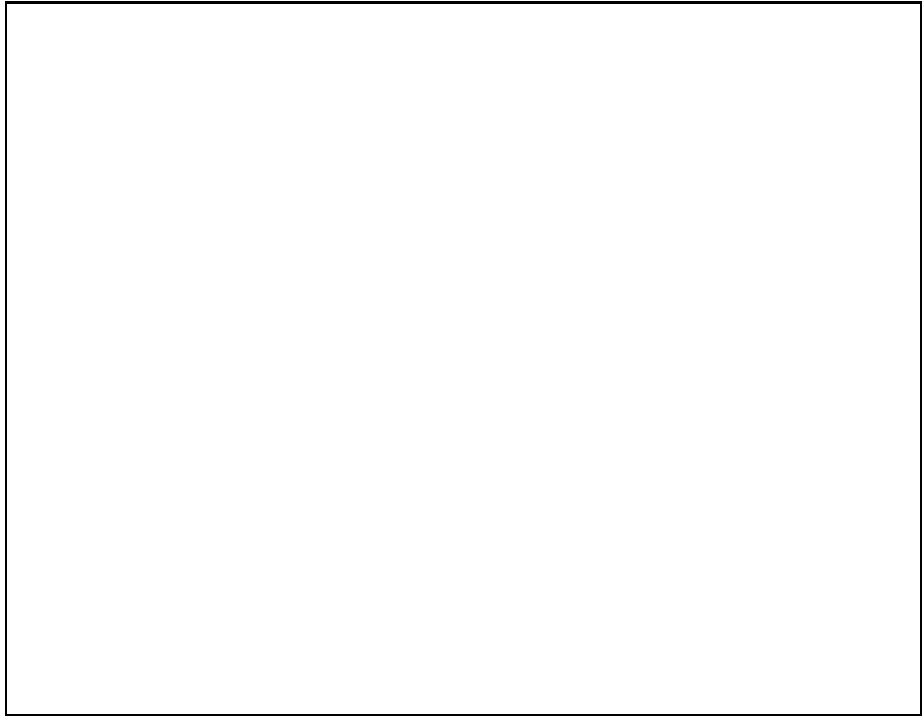
 
\picplace{9.5cm} 
\caption[]{B-band image of NGC 5128 shows the traditional, almost circular
inner part of the elliptical galaxy, crossed by a complex `dark band'.
The dark band is the projection of a strongly warped thin disk (Section
4). Note bright star clusters at the northwestern and southeastern
edges of the band, and isolated dust clouds silhouetted against
the stellar background. Supernova 1986g is the brightest of the two
stars seen in the dark band about 45 mm from the left and 40 mm from the
bottom. (Courtesy D. Malin, Ango-Australian Observatory)
}            
\end{figure} 

Other radio measurements soon established the nonthermal nature of 
the radio emission (Haddock, Mayer $\&$ Sloanaker 1954). Subsequent 
radio observations with increased sensitivity yielded ever-larger 
dimensions for the object until the low-frequency studies by Sheridan 
(1958) and Shain (1958) first established an overall size 
of 8$^{\rm o}$ $\times$ 4$^{\rm o}$ for Centaurus A -- much larger than 
the 20$'$ size of the optical parent galaxy NGC 5128. A multi-frequency 
radio study by Cooper, Price $\&$ Cole (1965) showed that at decimetre
wavelengths the source consists
of bright central emission, representing 20$\%$ of the total flux density 
and much larger lower-brightness lobes to the north and south, with 40$\%$ 
of the radio flux each. The advent of interferometric techniques, in 
particular aperture synthesis radio telescopes with their superior angular 
resolution, has since led to the discovery of many more similar radio 
galaxies ({\it cf.} Allen, Hanbury Brown $\&$ Palmer 1963; see review 
by Miley 1980). It has also become clear that many other elliptical galaxies 
have more or less pronounced dust lanes (Hawarden et al. 1981; Sadler $\&$ 
Gerhard 1985; van Dokkum $\&$ Franx 1995), including the other major southern 
radio galaxy NGC 1316 (Fornax A).

Both its striking optical appearance (Fig.~1), and its associated giant radio 
source have made NGC 5128\footnote{Useful photometric UBVR(I) 
sequences for NGC 5128 have been published by Graham (1980), McElroy $\&$ 
Humphreys (1982) and Zickgraf et al. (1990), reaching down to visual 
magnitudes of 16.6, 18.2 and 22.7 magnitudes respectively. Astrometric 
coordinates of field stars near the centre of NGC 5128 can be found in 
Griffin (1963). Very accurate field star positions and magnitudes may be 
obtained from http://astro.estec.esa.nl/SA-general/Projects/Hipparcos.}
into one of the most extensively studied galaxies in the southern 
hemisphere. However, in their comprehensive review Ebneter 
$\&$ Balick (1983) point out that NGC 5128 if placed at similar large distances
would look very much like other radio elliptical galaxies, in spite of 
its apparent peculiarity. By its fortuitous proximity, Centaurus A 
uniquely allows detailed studies
aimed at determining the nature of the 
galaxy and in particular the origin of the giant radio source it is hosting. 
A probably incomplete search of the literature published since Ebneter 
$\&$ Balick's (1983) review reveals over a hundred refereed papers mentioning 
Centaurus A in the title, whereas in the same period the number of papers 
referring to Centaurus A was almost 700. This review focusses on the observed 
properties of Centaurus A/NGC 5128 and their immediate interpretation. 
For a more complete review of the early work up to 1983, the reader is 
referred to Ebneter $\&$ Balick (1983), and for more general information to
recent proceedings such as `The Second Stromlo Symposium -- The Nature of
Elliptical Galaxies' (Eds. N. Arnoboldi, G.S. da Costa, P. Saha, 1997
PASP Conference Series Vol. 116), `Energy Transport in Radio Galaxies and
Quasars' (Eds. P.E. Hardee, A.H. Bridle, J.A. Zensus, 1996 PASP Conference
Series Vol. 100) and `Extragalactic Radio Sources' (Eds. R. Ekers, C. Fanti,
L. Padrielli, 1996 IAU Symposium 175).

\subsection{Distance}

With its great apparent brightness and galactocentric velocity of about
$V_{\rm o}$ = +325 $\kms$, NGC 5128 is clearly a nearby galaxy. Early 
estimates of its distance ranged from 2.1 Mpc (Sersic 1958) to 8.5 Mpc 
(Sandage $\&$ Tammann 1974) and a `compromise' distance of 5 Mpc has 
frequently been used in the past. Many of the earlier distance values 
were derived by indirect means: group membership, redshift, comparison to 
Virgo cluster distance (assuming the latter to be known) etc. Summaries of 
the various distance determinations prior to 1993 can be found in Shopbell, 
Bland-Hawthorn $\&$ Malin (1993) and especially de Vaucouleurs (1993).

Attempts to use the occurrence of the SNIa supernova 1986g in NGC 5128 to 
derive its distance (Frogel et al. 1987; Phillips et al. 1987; Ruiz-Lapente 
et al. 1991) were frustrated by uncertainties in the internal reddening of 
NGC 5128 and in the absolute magnitudes of the supernova (see Sect.~3.3). 
It appears more useful to {\it assume a distance} to determine the supernova
properties than the other way around. Other methods have been more succesful,
consistently yielding a relatively nearby distance. The accuracies of 
these techniques were reviewed by Jacoby et al. (1992). Tonry $\&$ Schechter 
(1990) interpret the globular cluster counts by Harris et al. (1984a, 1986) 
to imply a distance modulus $(m-M)_{\rm o}$ = 27.53$\pm$0.25 mag. Planetary 
nebula counts by Hui et al. (1993a) yield $(m-M)_{\rm o}$ = 27.73$\pm$0.14. 
The lower value $(m-M)_{\rm o}$ = 27.48$\pm$0.06 obtained by Tonry $\&$ 
Schechter (1990) from the surface brightness fluctuation of globular clusters 
must be revised to $(m-M)_{\rm o}$ = 27.71$\pm$0.10 (Tonry 1991). 
Finally, Soria et al. (1996) estimate $(m-M)_{\rm o}$ = 27.72$\pm$0.20 from 
halo red giant branch stars observed with HST/WFPC2. As there is no a priori 
preference for any of these methods, and as the weighted and unweighted 
averages of the various derived distance moduli are practically identical, 
the best value is thus $(m-M)_{\rm o}$ = 27.67$\pm$0.10, corresponding to a 
distance of $D$ = 3.4$\pm$0.15 Mpc. Note that at this distance, 1$'$ on the 
sky corresponds almost precisely to 1 kpc.\footnote{In the following all 
specific values quoted are reduced to $D$ = 3.4 Mpc so that they may be 
different from the published values as given in the associated reference.}

\subsection{Systemic velocity}

The systemic velocity of NGC 5128 is most frequently derived from measurements 
of components of the disk (Sect.~4). An overview of estimates up to 1984 can
be found in Hesser et al. (1984); Table 1 lists all estimates {\bf a.} later 
than 1975; {\bf b.} having a quoted accuracy of 10 $\kms$ or better; and 
{\bf c.} {\it not} based on HI or molecular line absorption (see Sect.~7).
The estimates range from 536 to 551 $\kms$, and the mean is $V_{\rm Hel}$ = 
543$\pm$2 $\kms$.
 
\begin{table}[htb]
\centering
\caption[]{Systemic velocity estimates }
\begin{tabular}{lll} 
\hline\noalign{\smallskip}
Systemic Velocity		&  Obtained from:	& Reference \\
$V_{\rm Hel}$ ($\kms$)   	&			& \\
\noalign{\smallskip}
\hline\noalign{\smallskip}
548$\pm$5	& nebular emission lines & Graham (1979) \\
551.4		& H$\alpha$ emission	 & Whiteoak $\&$ Gardner (1979) \\
541$\pm$8	& nebular emission lines & Rodgers $\&$ Harding (1980) \\
545$\pm$5	& H$\alpha$ emission	 & Marcelin et al. (1982) \\
538$\pm$10  	& stellar absorption lines & Wilkinson et al. (1986) \\
536$\pm$5	& H$\alpha$ emission 	 & Bland et al. (1987) \\
542$\pm$7   	& HI emission 		 & van Gorkom et al (1992) \\
541$\pm$5	& CO modelling		 & Quillen et al. (1992) \\
541$\pm$7 	& PN emission lines   	 & Hui et al. (1995) \\
\noalign{\smallskip}
\hline
\end{tabular}
\end{table}

\begin{figure}[t]
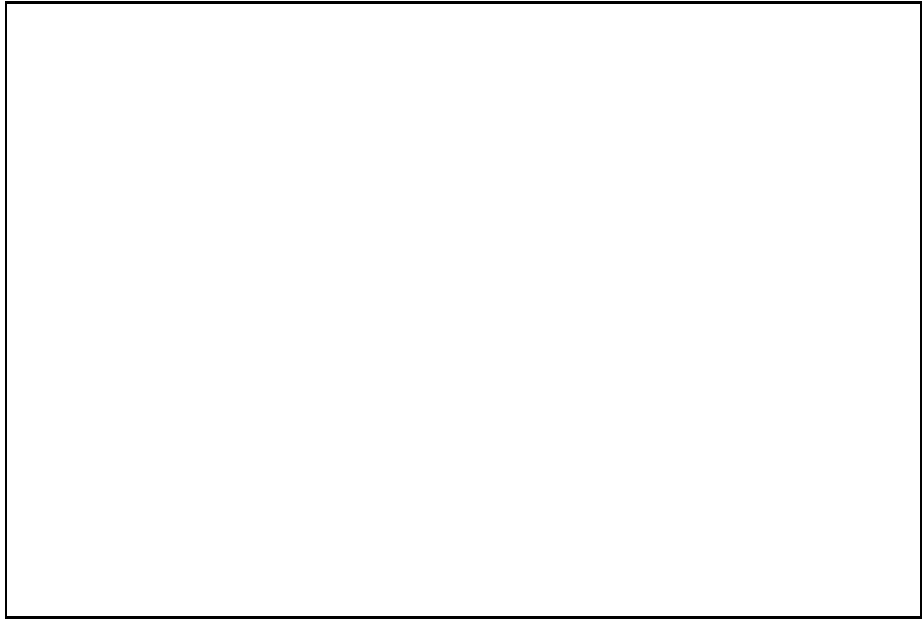
 
\picplace{8.2cm} 
\caption[]{The Centaurus group of galaxies, with its brighter members
identified. From C\^ot\'e et al. (1997).
}            
\end{figure} 

\subsection{Environment}

Based on position and radial velocity, NGC 5128 is part of a group 
of 25 galaxies extending over about 25$^{\rm o}$ on the sky (de Vaucouleurs 
1975; Webster et al. 1979; Hesser et al. 1984; C\^ot\'e 1995, 1997). In 
addition to NGC 5128, major members of the group are NGC 4945 and NGC 5236 
(M 83) as well as the lesser NGC 5102, NGC 5253 and some twenty dwarf 
galaxies (Fig.~2). NGC 5128 is the only massive elliptical in the group.
The group galactocentric velocity is very close to the 
individual velocities $V_{\rm o}$ = 320--330 $\kms$ of the major members NGC 
4945, NGC 5128 and NGC 5236 (Hesser et al. 1984). The members have an average 
projected radial distance to the centre of mass of 0.72 Mpc. The group has 
a crossing time of 5 gigayear and is probably still collapsing (C\^ot\'e 1995, 
1997). Assuming a distance of 3.4 Mpc, the group mass estimate by Hesser 
et al. (1984) becomes 5--17 $\times$ 10$^{12}$ $\Msun$. It is quite remarkable 
that the other two major members, NGC 4945 and NGC 5236, as well as NGC 5253, 
exhibit signs of unusually vigorous star formation, while NGC 4945 also has a 
nuclear outflow. In addition, the dwarfs in the group have relatively high
central surface brightnesses and exhibit clear star formation activity 
(C\^ot\'e 1995).

NGC 5128, itself a strong X-ray source (Sects. 2.4; 5.5), is surrounded 
by several X-ray point sources of a nature still unknown, although the 
majority of them seems to be associated with the galaxy (Arp 1994; 
D\"obereiner et al. 1996).

\section{The radio source and associated features}

\subsection{Overview}

Owing to its proximity, Centaurus A (PKS 1322-427) is the largest 
extragalactic radio source in the sky. It extends predominantly in a 
north-south direction between declinations $-38.5^{\circ} \leq \delta 
\leq -46.5^{\circ}$, and between right ascensions 13$^{h}$15$^{m} \leq 
\alpha \leq$ 13$^{h}$32$^{m}$ (Fig.~3). Its overall angular dimensions are 
8$^{\circ} \times$ 4$^{\circ}$ (Cooper et al. 1965; Haynes et al. 1983; 
Junkes et al. 1993; Combi $\&$ Romero 1997). This translates into a linear 
size of 500\,$\times$\, 250 kpc. Although larger radio galaxies have been 
found ({\it cf.} Miley 1980), Centaurus A is still, in an 
absolute sense, one of the largest known. Its moderate radio luminosity 
places it in class I of the classification by Fanaroff $\&$ Riley (1974),
suggesting only moderate relavistic beaming effects (Jones et al. 1993).
For a recent study of the properties distinguishing FR-I and FR-II radio 
galaxies, see Zirbel (1996), and references therein.

The radio source is very complex; it shows significant structure ranging over 
a factor of 10$^{8}$ in size from the largest scales down to less than a 
milli-arcsecond.
Major components of the radio source (see Fig.~3) are
the giant outer lobes extending to 250 kpc, the northern middle 
lobe (no southern counterpart) extending to about 30 kpc, 
the inner lobes and central jets extending to about 5 and 1.35 kpc 
respectively and the compact core with associated 
nuclear jets extending over 1 pc (see Fig.~11).

\begin{figure}[t]
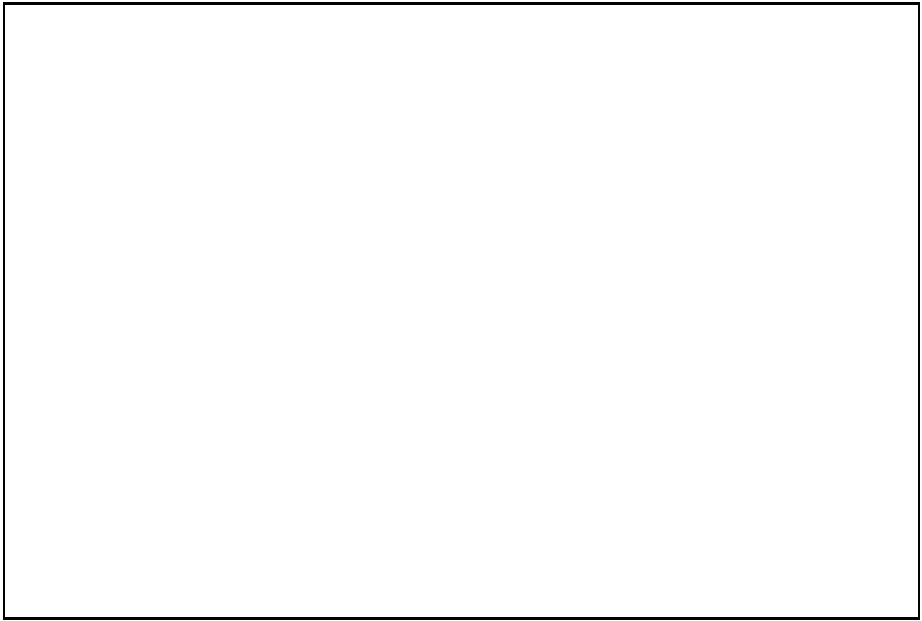
 
\picplace{8.2cm} 
\caption[]{Radio maps of Centaurus A, highlighting the various components
of the radio source introduced in Sect.~2.1. From Burns et al. (1983).
}            
\end{figure} 

Table 2 provides a guide to radio continuum observations of Centaurus A. 
Part A lists radio continuum maps covering the {\it whole} source; 
integrated flux-densities are not always provided by the authors. Total source 
flux-densities between 85 MHz and 4750 MHz define an overall spectral index 
$\alpha$ = $-0.6$\footnote{Spectral index $\alpha$ is defined by 
$S_{\nu} \propto \nu ^{\alpha}$}. Part B refers to aperture synthesis radio 
maps of the bright inner part which comprises the inner lobes, jets and core, 
and single-dish maps at very high frequencies. References to other early 
single-dish maps and flux-densities at various frequencies can be found in 
Slee et al. (1983) and Tateyama $\&$ Strauss (1994). Part C, finally, 
summarizes radio images and flux-densities of the nuclear source. Very 
high resolution (VLBI) observations have shown a complex structure even 
for the inner few parsec which is further discussed in Sect.~5.4. Here, 
we draw attention to the high-frequency variability of the source 
which is clear from the Table and which is further discussed in Sect.~5.6.

\begin{table}[htbp]
\centering
\caption[]{Radio continuum observations of Centaurus A}
\begin{tabular}{lrcr} 
\hline\noalign{\smallskip}
Reference		 & Frequency	& Resolution &  Total Flux \\
			 & (MHz)	&            &  (Jy) \\
\noalign{\smallskip}
\hline\noalign{\smallskip}
\multicolumn{4}{c}{A. Whole Source} \\
Shain (1958)		 &    20	&  		&  28\,000 \\
Sheridan (1958)		 &    85	&		&  $^{a}$7400 \\
Cooper et al. (1965)	 &   406	&   48$'$	&  2710	\\
Haslam et al. (1981)     &   408	&   48$'$	&  \\
Bolton $\&$ Clark (1960) &   960	&   20$'$	&  $^{a}$1675 \\
Cooper et al. (1965)	 &  1410	&   14$'$	&  1330 \\
Combi $\&$ Romero (1997) &  1435	&   30$'$	&  \\
Cooper et al. (1965)	 &  2650	&   7.4$'$	&  \\
Junkes et al. (1993)	 &  4750	&   4.3$'$ 	&  681 \\
Haynes et al. (1983)	 &  5000	&   4.1$'$ 	&  \\
\noalign{\smallskip}
\hline\noalign{\smallskip}
\multicolumn{4}{c}{B. Inner Lobes $\&$ Core} \\
\noalign{\smallskip}
\hline\noalign{\smallskip}
Slee et al. (1983)	&    327	&  55$''$ 		  & 734 \\
Slee et al. (1983)	&    843	&  40$'' \times$60$''$	  & 392 \\
Christiansen et al. (1977) & 1415	&  50$'' \times$45$''$	  & 215 \\
Clarke et al. (1992)	&   1446	& 4.5$'' \times$1.2$''$   & 78.3 \\
Clarke et al. (1992)	&   1634	& 4.5$'' \times$1.2$''$   & 63.5 \\
Clarke et al. (1992)	&   4866	& 4.5$'' \times$1.2$''$   & 37.5 \\
Tateyama $\&$ Strauss (1992) & 22\,000	& 258$''$		  & 51.2 \\
Tateyama $\&$ Strauss (1992) & 43\,000	& 132$''$		  & 31.2 \\
\noalign{\smallskip}
\hline\noalign{\smallskip}
\multicolumn{4}{c}{C. Nucleus $\&$ Nuclear Jet} \\
\noalign{\smallskip}
\hline\noalign{\smallskip}
Slee et al. (1983)	&    327	&  55$''$ 		  &  2: \\
Slee et al. (1983)	&    843	&  40$'' \times$60$''$	  &  3: \\
Schreier et al. (1981)	&   1407	& 31$'' \times$9.5$''$	  & 3.4 \\
Clarke et al. (1992)	&   1446	& 4.5$'' \times$1.2$''$   & 3.7 \\
Clarke et al. (1992)	&   1634	& 4.5$'' \times$1.2$''$   & 4.9 \\
Jones et al. (1994)	&   2300	& 0.1$''$		  & 3.4 \\
Schreier et al. (1981)	&   4866	& 31$'' \times$9.5$''$	  & 5.1 \\
Clarke et al. (1992)	&   4866	& 1.1$'' \times$0.3$''$   & 6.9 \\
Jones et al. (1994)	&   8400	& 0.03$''$		  & 5.9 \\
Clarke et al. (1992)	&  14984	& 0.44$'' \times$0.14$''$ & 9.4 \\
Botti $\&$ Abraham (1993) & 22\,000	& 252$''$		  & $^{b}$16--32 \\
Fogarty $\&$ Schuch (1975) & 22\,000	& 252$''$		  & 21.5 \\
Kellerman (1973)	&  31\,500	& 225$''$		  & 24 \\
Tateyama $\&$ Strauss (1992) & 43\,000	& 132$''$		  & 10.8 \\
Botti $\&$ Abraham (1993) & 43\,000	& 132$''$		  & $^{b}$7--20 \\
Kellerman (1973)	& 89\,000	&  80$''$		  & 18 \\
Israel (unpublished)    & 89\,000	&  57$''$		  & $^{c}$8 \\
Israel (unpublished)	& 110\,000	&  47$''$		  & $^{c}$7 \\
Israel (unpublished)	& 145\,000	&  36$''$		  & $^{c}$6 \\
\noalign{\smallskip}
\hline
\end{tabular}

Notes: a. As revised by Cooper et al. (1965); b. Range of variation over period
1979 -- 1992; c. Measurements in September 1996. \\

\end{table}

\begin{figure}
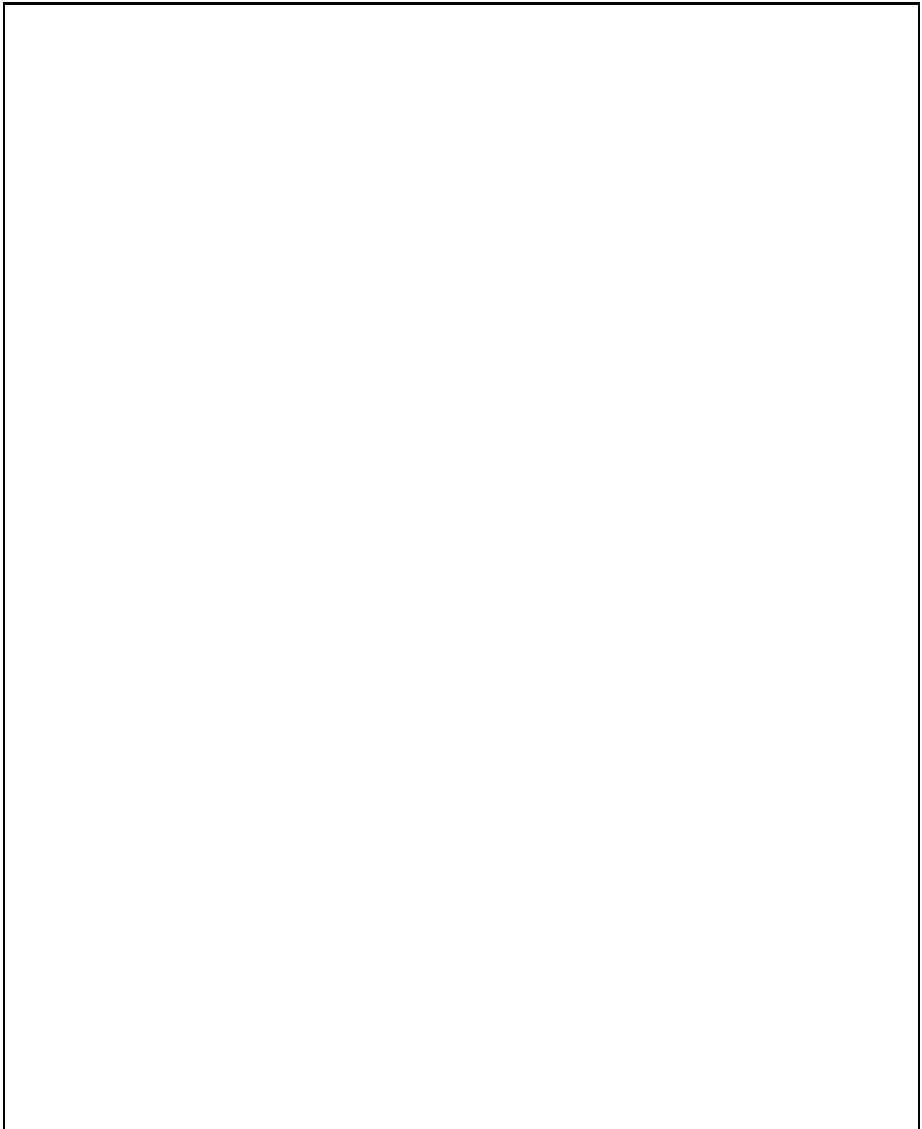
 
\picplace{15.0cm} 
\caption[]{Radio map of Centaurus A at 4.75 GHz. Contours reach from
50 to 40\,000 mJy/beam area; HPBW beam size is 4.3$'$. E-vectors,
marked for polarized intensities between 25 and 250 mJy/beam area, indicate 
both intensity and direction of polarization. The map shows the extent of
the outer and middle lobes discussed in Sect.~2.2. The inner lobes and
jets, discussed in Sects.~2.3 and 2.4 are represented by the barely
resolved strong double source in the centre. From Junkes et al. (1993).
}            
\end{figure} 

The whole radio source has been mapped with single-dish telescopes only, at 
linear resolutions no better than 4 kpc. The radio-bright inner lobes have 
been mapped with aperture synthesis telescopes at hundredfold higher 
resolutions of about 40 pc and the inner jets and core have been mapped 
at (sub)parsec resolutions in VLBI experiments. Aperture synthesis mosaicing 
of the outer features at substantially higher linear resolutions should be 
undertaken. Although a daunting task in view of the large angular sizes and 
low surface brightnesses, it is of great importance as no other radio galaxy 
allows its lobe structures to be studied in the detail possible in the case 
of Centaurus A. 

\subsection{The outer and middle lobes}

The overall distribution of radio emission is roughly $S$-shaped. The  
distribution of radio emission across the {\it northern and southern outer 
lobes} is very asymmetric. The brightest part of the giant northern lobe is 
usually called the northern middle lobe; there is no such feature in 
the giant southern lobe for reasons that are not understood. One of the dwarf 
companion galaxies, UKS 1324-412 = ESO 324-G024, is seen in projection against 
the centre of the giant northern lobe. The isolated radio peak at 
$\alpha$ = $13^{h}18{m}$, $\delta$ = $-43^{\circ}26'$ (Fig.~4) is not part 
of the southern lobe, but is due to a 15 mag background elliptical radio 
galaxy (Cooper et al. 1975; Haynes et al. 1983). This galaxy
could be used as a tracer for conditions in the lobe (Junkes et al. 1993). 
Another unrelated dwarf galaxy, somewhat misleadingly known as the 
Fourcade-Figueroa Shred, is located
at the tip of the southern main lobe (Dottori $\&$ Fourcade 1973). While 
radio surface brightnesses steadily decline away from the core in the northern 
lobe, they reach a maximum in the southern lobe at 1.8$^{\circ}$ from the core.
The position angle of the northern lobe is close to 0$^{\circ}$, 
while the southern outer lobe is displaced to the west and has a position 
angle of roughly 135$^{\circ}$. The two lobes are connected by a low surface 
brightness `bridge'  and the northern lobe is probably closer to us than the 
southern lobe (Junkes et al. 1993). Most of the emission from the lobes has a 
spectral index $-0.5 \leq \alpha \leq -0.7$, characterized by a ratio of 
random to uniform magnetic field strength $B_{\rm r}/B_{\rm o}$ = 0.6 (Combi 
$\&$ Romero 1997). The outer lobes may be associated with hard X-ray limb 
brightening (Arp 1994).

The {\it northern middle lobe} is a major feature of the radio source: at 5 
GHz it contributes almost 45\% to the total radio emission from Centaurus A. 
Its overall orientation is north-south at $13^{h}23^{m}.5$ and it fades into 
the outer lobe at about $\delta$ = $-41^{\circ}30'$. Its 
radio peak is at a projected distance of 20 kpc to the nucleus, at position 
angle 36$^{\circ}$ (Haynes et al. 1983). Whereas its radio intensities 
drop rather gently towards the north and northwest, its southeastern edge is 
relatively sharply delineated. The middle lobe is associated with soft X-ray 
emission, with a luminosity of about 5 $\times$ 10$^{39}$ erg s$^{-1}$
(Feigelson et al. 1981). Although inverse Compton scattering of the microwave 
background a priori is expected to occur in extended radio lobes, this 
explanation does not seem to fit the X-ray emission from this lobe (Feigelson 
et al. 1981; Marshall $\&$ Clark 1981; Morini et al. 1989). High-resolution 
radio observations -- not yet existing -- might enable a choice to be made 
between the remaining explanations of thermal and synchrotron emission.

The polarization E-vectors (perpendicular to the magnetic field direction) 
are well-aligned, but change their direction with increasing right ascension
across the middle lobe from position angle $\approx$ 15$^{\circ}$ to 
90$^{\circ}$, emphasizing its concave southeastern boundary (Fig.~4; Junkes 
et al. (1993). Polarized intensities are between 20\% and 40\%, and the 
polarization from the middle lobe continues smoothly into that of the 
northern outer lobe where peak polarized intensities reach somewhat higher 
levels of up to 50\% . The southern outer lobe has polarized intensity levels 
similar to those in the north, but south of $\delta$ = $-44^{\circ}30'$ its 
polarization structure abruptly becomes very chaotic. This could be due either 
to turbulence in the southern lobe, or to the presence of Galactic foreground 
polarization. Junkes et al. (1993) entertained the possibility that the giant 
northern lobe is bending into our line of sight, while the southern lobe may 
be turning away from the plane of the sky in the opposite direction.

In both lobes, steeper spectra are found where the radio surface brightness 
declines to low levels (see Fig.~4). These steep-spectrum regions appear to 
be unpolarized, but this may merely reflect a lack of sensitivity (Combi $\&$ 
Romero 1997; Junkes et al. 1993). 

\subsection{The inner lobes and outer jet}

The two inner lobes ({\it cf.} Fig.~3) together contribute almost 30\% to the 
total 5 GHz flux-density of Centaurus A. The radio emission from the northern 
inner lobe is about 40\% higher than that from the southern inner lobe. Slee 
et al. (1983) show that both lobes have identical nonthermal spectral indices 
$\alpha$ = $-0.7$, and estimate from this an age of 6 $\times$ 10$^{8}$ years. 

\begin{figure}[t]
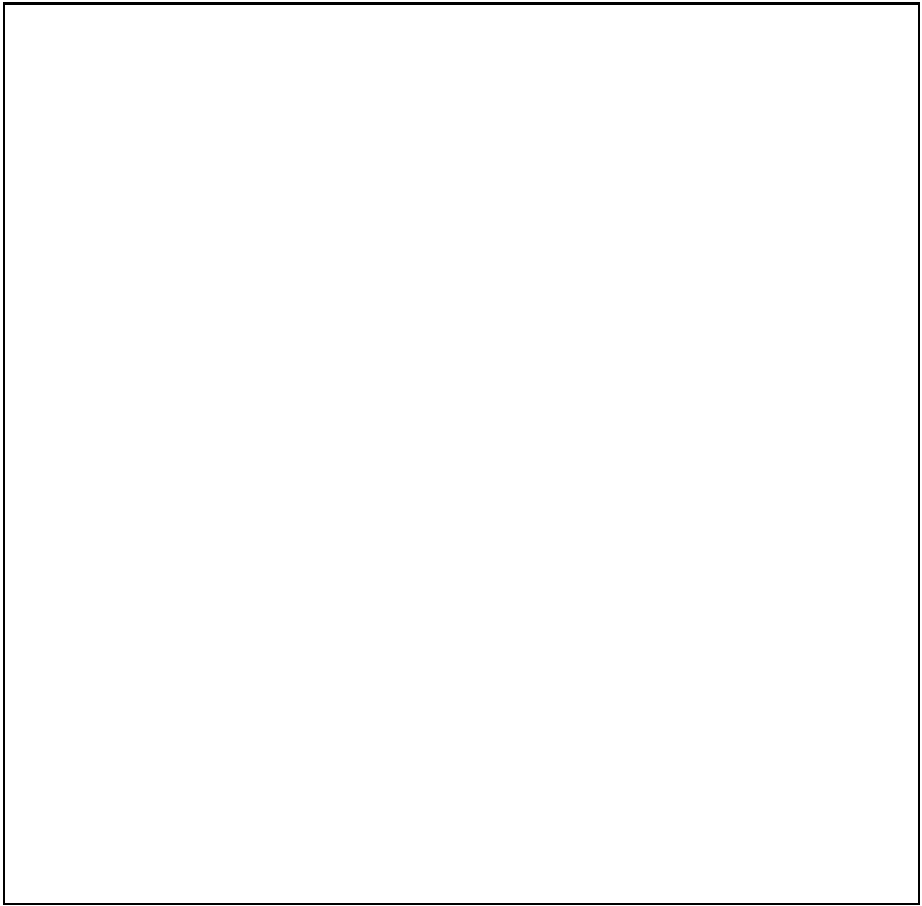
 
\picplace{12.0cm} 
\caption[]{Gray scale 5 GHz image of the inner lobes and jets of Centaurus
A at a resolution of 4.4$\times$1.2 arcsec. The gray scales range linearly
from -3.0 mJy per beam (dark) to 35.0 mJy per beam (light). The labelled
features are discussed in Clarke et al. (1992). Courtesy J. Burns, Univ.
of Missouri-Columbia).
}            
\end{figure} 

With a maximum angular size of 12$'$, the inner lobes are usually not, or 
poorly, resolved by single-dish telescopes. Aperture synthesis maps (Fig.~5)
reveal pronounced filamentary structure especially in the southern lobe,
probably generated by velocity shear in the turbulent flow of the lobe rather 
than by shocks (Clarke et al. 1992). The northern lobe has a fairly 
sharply delineated southeastern edge. At the same position angle as the inner 
lobes, 
Arp (1994) has found a jet of diffuse X-ray emission, extending, however, much 
beyond northern radio lobe limits. Its nature and origin are not clear. No 
soft X-ray emission is seen towards the inner lobes (Feigelson et al. (1981).

In the high-resolution maps shown by Clarke et al. (1992), the inner 
lobes are revealed as subsonic plumes emanating from a supersonic central jet 
at a position angle of 51$^{\circ}$. The jet--lobe transition occurs at the 
location of the innermost optical shell identified by Malin et al. (1983) 
and marked No. 9 in Fig.~7. This transition has been interpreted as 
evidence for shock disruption of the jet at the interface of the interstellar 
and intergalactic medium of the galaxy (Norman et al. 1988; Gopal-Krishna 
$\&$ Saripalli 1984). 

The northern jet is quite pronounced and shows several knots, whereas the
southern (counter)jet was first surmised from the detection of a few weak 
knots in the southern lobe (Clarke et al. 1986; 1992); it is now confirmed 
by VLBI images (Sect.~5.4). The direction of polarization in the inner lobes 
is dramatically different from that in the middle and outer lobes. Beyond the 
sharply delineated northern plume perimeter, at a projected distance of 6.3 
kpc to the nucleus, it changes abruptly by 90$^{\circ}$, so that the northern 
inner lobe is quite distinct from the middle lobe. E-vector alignments 
indicate a magnetic field emanating radially from the nucleus along the lobe 
axis; at the edges of both lobes, the magnetic field is parallel to the plume 
perimeter (Clarke et al. 1992). At 5 GHz, both lobes are $\approx$ 35\% 
polarized with some regions having polarizations as low as 15\% or as high as 
50\%. The southern lobe is more strongly depolarized, most likely because it 
is located behind the clumpy interstellar medium of the galaxy's disk 
component (Sect.~4). It exhibits a much less ordered polarization pattern 
(Clarke et al. 1992; Junkes et al. 1993). 

Broad-line high excitation gas appears to be associated with the inner lobes,
but its nature is still unclear (Phillips et al. 1984 and references therein).

\subsection{Inner jet structure}

The morphologies of the inner radio jet (Schreier et al. 1981) and the X-ray 
jet (Schreier et al. 1979; Feigelson et al. 1981; D\"obereiner et al. 1996; 
Turner et al. 1997) are very similar. They fill the gap between the
nucleus and the optical filaments reported by Dufour $\&$ van den Bergh
(1978 -- Sect.~2.5). Seven major concentrations (usually labelled A through
G following Feigelson et al. 1981) can be discerned with rather identical
radio/X-ray flux ratios. The knots are brightest nearest the nucleus and are 
separated by typically 30$''$. They have 0.4--4.5 keV X-ray luminosities of 
0.2--1.0 $\times$ 10$^{39}$ erg s$^{-1}$. Once again, it appears that inverse 
Compton radiation cannot explain their nature, although both thermal and 
synchrotron models are possible (Feigelson et al. 1981; see also Turner et al. 
1997). The kinetic energy of the jet averaged over time exceeds the radiative
energy of the nucleus. It has been suggested that both radio and X-ray
emission are produced by a single population of relativistic electrons with 
Lorentz factors $\gamma \approx$ 8 $\times$ 10$^{7}$ in 60 microgauss magnetic 
fields (Schreier et al. 1981; Burns et al. 1983). As such electrons have 
lifetimes of 50 years or less, in any event much shorter than light travel 
times {\it even across the knots}, this requires in situ acceleration 
(Schreier et al. 1981; Feigelson et al. 1981; Burns et al. 1983; Turner et al. 
1997). There is no evidence for X-ray flux variations of the {\it inner jet} 
(D\"obereiner et al. 1996; Turner et al. 1997), in contrast to the {\it
nuclear jet} (see Sect.~5.6).

High resolution radio images (18$\times$5 pc) by Clarke et al. (1986; 1992) 
illustrate the extremely inhomogeneous nature of the main (northern) jet, 
with a hierarchy of knots reaching down to sizes less than 2.5 pc. The jet 
is not a linear structure, but exhibits side-to-side limb brightening, 
perhaps due to a combination of external and internal shocks (Clarke et al. 
1986). As most of the jet is inside the main body of the galaxy, some degree 
of interaction with the gaseous medium of the galaxy is to be expected. 
Indeed, Burns et al. (1983) suggested that the inner jet is confined by 
external pressure. In the knots, the spectral index decreases from 
$\alpha = -0.7$ at the limb-brightened side to $\alpha = -1.4$ at the 
diffuse side, while polarized fractions drop from 40\% to less than 20\% . 
Filaments or streamers connect knot complexes to ones downstream; they have 
steep spectra $\alpha = -1.4$. If the knots are shock fronts reaccelerating 
electrons, the streamers will represent material flowing down the jet away 
from these fronts at speeds of 5000 $\kms$; the spectral steepening reflects
increasing importance of synchrotron energy losses (Burns et al. 1983; Clarke 
et al. 1986).

\subsection{Optical jet features}

Intense optical emission from the knots in the brightest part of the 
X-ray/radio jet, at 0.25$'$--1.0$'$, and also from in between the knots, was 
discovered by Brodie et al. (1983). The emission is due to gas at densities of 
a few hundred per cc in the knots, and from gas at less than 10 per cc in the 
interknot region. At least one of the optical features corresponds precisely 
to radio/X-ray structure suggesting shock excitation of clumps carried along 
in the jet flow. The continuum appears to be the optical counterpart of the 
synchrotron X-ray/radio emission (Brodie $\&$ Bowyer 1985). 

Somewhat farther out, a long chain of filaments, dust clouds, H$\alpha$ knots 
and blue stellar objects emanates from the centre of the galaxy more or less 
along the direction of the jet. This chain starts at about 2$'$ from the 
nucleus and extends out to 28$'$ (see the review by Danziger 1981 and 
references therein). It breaks up into three distinct groups of features.

The socalled {\it inner jet structures} of this chain
are found between 2$'$ and 6$'$ (Dufour $\&$ van den Bergh 1978); the 
features closest to the nucleus are aligned with but offset from the radiojet, 
just where this starts to develop in the plume of the northern inner lobe. 
The outermost features of this group continue at the same position angle, 
although the radio plume expands away to the north. In their Fig.~1,
Morganti et al. (1991) provided high-quality images as well as a very useful
diagram of the relative positions of optical and radio structures. 
The next group out (commonly called the {\it inner filaments}) occurs between
8$'$ and 10$'$ from the nucleus at the same position angle as the jet
($PA$ = 55$^{\circ}$), although the outermost may start `veering off course'
(see Morganti et al. 1992). They are easily visible on long-exposure images 
of NGC 5128 ({\it cf.} Figs.~6 and 7). These filaments are in the northern middle 
radio lobe, but have no known specific radio counterpart. This may reflect 
the lack of high-resolution radio images of the middle lobe more than 
anything else and provides another rationale for obtaining such. The filaments 
have very high [OIII]/H$\beta$ ratios of about 15 and are blueshifted by up to 
400 $\kms$ from the systemic velocity of the galaxy (Osmer 1978; Morganti 
et al. 1991). They consist of turbulent low-density gas and coincide with 
the soft X-ray emission associated with the northern middle lobe 
(Feigelson et al. 1981; Graham $\&$ Price 1981; Graham 1983).

The {\it outer filaments} (at 17.5$'$) are also in the middle radio lobe,
again without obvious radio counterpart at least at the low resolutions
presently available. Their position angle, $PA$ = 43$^{\circ}$ is markedly 
different from that of the other filaments (Fig.~6). In addition to the 
brightest main filaments, there are weaker features to the north and west. 
Graham (1983) measured some rather high blueshifts up to 1000 $\kms$, but 
Morganti et al. (1991) found blueshifts of only 70--215 $\kms$ for the main 
filaments.

The origin and excitation of the filaments has been a matter of some debate.
Although some of the features have spectra similar to those of `normal'
HII regions, several (notably the inner filaments) have spectra suggesting
photo-ionization by a (nuclear) power-law spectrum or by shock excitation. 
Metallicities appear to rule out a primordial origin of the gas. The filaments 
may be formed by the expanding radio lobes compressing ambient gas to 
temperatures of a few million K followed by rapid cooling on a timescale of 
10 million years (Feigelson et al. 1981; Gouveia dal Pino $\&$ Opher 1989).
Alternatively, the jet may consist of material ejected from the nuclear region 
(Graham $\&$ Price 1981), although it is questionable whether this would 
survive long enough to have reached the presently observed distance to the 
nucleus (Graham 1983). In a rather extensive study of the filaments, Morganti 
et al. (1991, 1992) concluded that they are predominantly photoionized by the 
radiation from the nucleus itself, extinguished in our line of sight but 
travelling unimpededly along the jet. They proposed that a BL Lac nucleus 
lies at the core of Centaurus A. This conclusion was tentatively supported 
by Schiminovich et al. (1994) who found that the apparently counterclockwise 
rotating HI clouds near these filaments have velocities very similar to those 
found optically. However, Viegas $\&$ Prieto (1992) argued that additional 
heating mechanisms are required, and suggested the presence of shocks. They 
also concluded that a mixture of optically thin and optically thick gas is 
necessary for photoionization to be a viable explanation. In contrast, 
Sutherland et al. (1993) proposed that instead of photo-excitation, the 
mechanical flux of a mildly supersonic low-density jet interacting with
dense ambient clouds is sufficient to energize shock waves with velocities
$\approx$ 200--450 $\kms$. Their model reproduces several observed features
of the filaments, and removes the need for a narrow ionizing beam as 
proposed by Morganti et al. (1991, 1992). As the latter model predicts 
strong UV line fluxes from the filaments, whereas the former does not, 
such an observational test would be interesting.

Finally, we note that the inner filaments are at the same position angle
as the inner jets, the more distant outer filaments are at the same
position angle as the bright NL-2 component (Clarke et al. 1992; see
Fig.~5) in the plume, and the even more distant faint galaxy extensions 
in Fig.~8 line up with the NL-4 component in the plume (see Fig.~ 5). 
Whether this is accidental, or provides clues to their origin is unclear.

\section{The galaxy} 

\subsection{Overview}

With $V$ = 6.98 mag (de Vaucouleurs et al., 1976 -- RC2), 
NGC 5128 is the fifth brightest galaxy in the sky,
immediately after the Local Group members M 31, M 33, LMC and SMC. Images
of relatively short exposure, limited to a surface brightness in B of about 
22 mag per arcsec$^{2}$, show an almost circular appearance, which has led to 
the classification S0p or E0p (Fig.~1). However, at lower surface-brightness 
levels (Fig.~8), the shape of the galaxy becomes increasingly noncircular. At 
about 25 mag per arcsec$^{2}$, the axial ratio has increased to 1.3, after 
which elliptical symmetry is lost ({\it cf.} Haynes et al. 1983). A more 
appropriate classification appears to be E2 (Dufour et al. 1979; McElroy $\&$ 
Humphreys 1982; Haynes et al. 1983; Ebneter $\&$ Balick 1983), with the 
photometric major axis of the galaxy at position angle 35$^{\circ}$. The light 
distribution closely follows the $r^{1/4}$ de Vaucouleurs law (van den Bergh 
1976) characteristic of elliptical galaxies. The inner parts of NGC 5128 have 
a roughly constant total mass-to-light ratio inccreasing with radius, probably
due to the presence of dark matter (Sect.~6.2).

\begin{figure}[t]
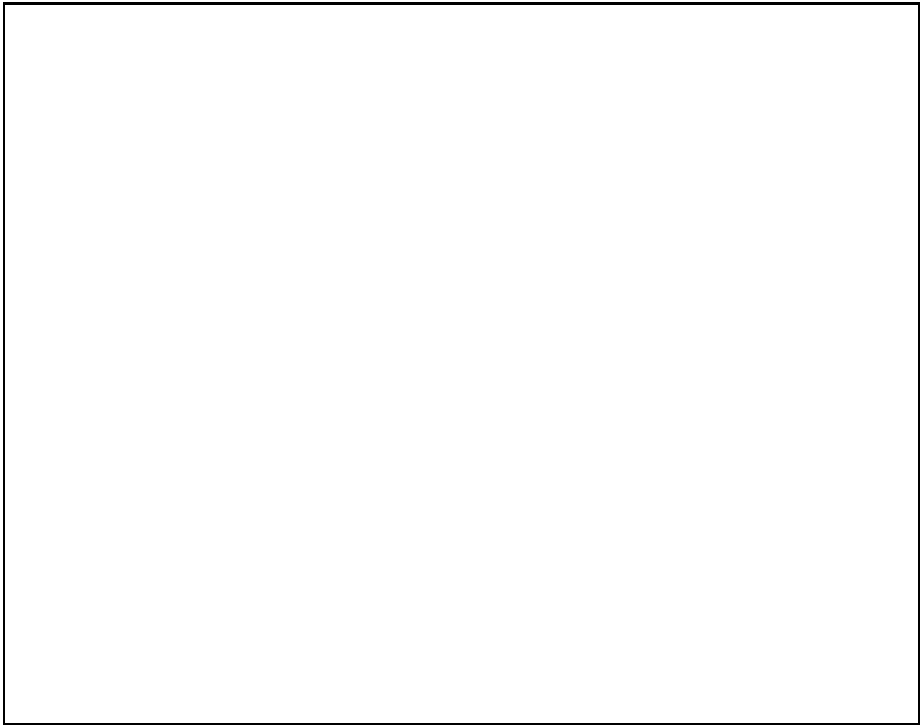
 
\picplace{9.6cm} 
\caption[]{Deep negative B-band image of NGC 5128 shows the system of
optical shells at the edges of the galaxy. This system of shells is one
of the clearest indications that NGC 5128 has undergone mergers in its
past. Also note optical jet features at a position angle of about 45$^{\circ}$,
inside and outside the northeastern shells respectively. 
Courtesy D. Malin, Ango-Australian Observatory)
}            
\end{figure} 

Of particular interest are the deep and specially processed optical images 
presented by Malin et al. (1983 -- see Fig.~6) and Haynes et al. (1983 -- see
Fig.~8). 
The former reveals an extensive system of shells of old disk stars within the 
extended elliptical galaxy. The system is most regular on the northeastern 
side, and more fragmented on the southwestern side. The outermost shell is 
found at 18$'$ from the nucleus. Atomic neutral hydrogen emission with a total
mass of 1.5$\times$10$^{8}$ $\Msun$ is present just outside several of the 
outer shells (Schiminovich et al. 1994), with a rotation axis at position 
angle 285$^{\circ}$, somewhat offset from the position angle of the minor 
axis of the E2 galaxy and the dust band (Fig.~7). 

\begin{figure}[t]
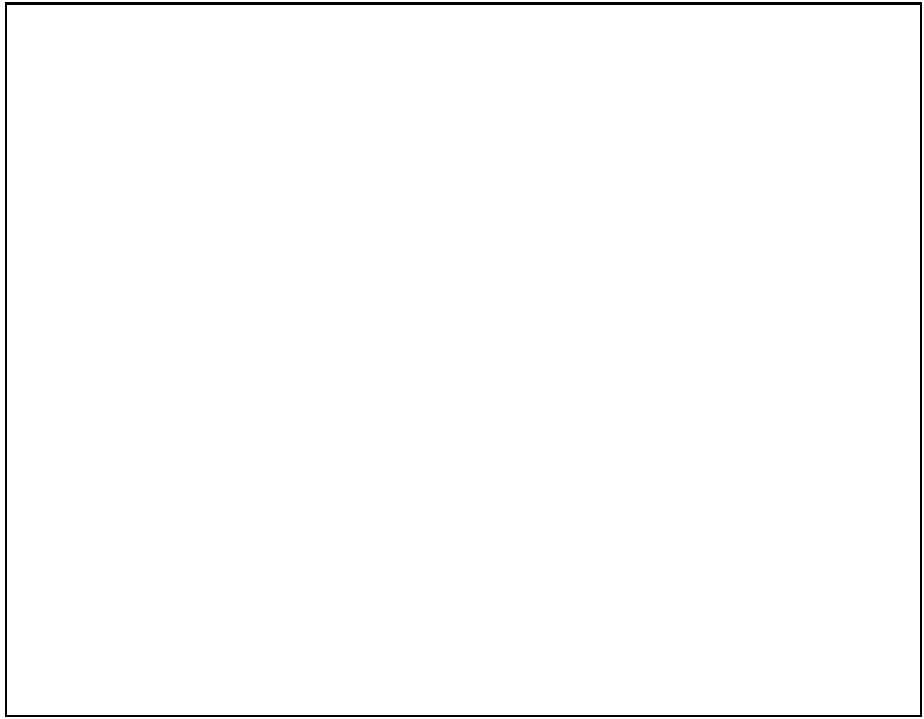
 
\picplace{9.5cm} 
\caption[]{Contours of total HI superposed on a schematic image of the
shells of NGC 5128. Contours reach from 1 to 40 $\times \twcm$.
Thick lines mark the position of sharp (solid) and diffuse (dashed) shells.
Shaded regions represent dust patches. Thin contours marked A$_{n}$ and
A$_{s}$ represent the radio continuum emission from the inner lobes and
jets discussed in Sects.~2.3 and 2.4. Most of the neutral hydrogen is 
associated with the warped disk discussed in Sect.~4, but some of it
is associated with the shells (Sect.~3.1). From Schiminovich et al. 1994;
image courtesy J.M. van der Hulst.
}            
\end{figure} 

The image presented by Haynes et al (1983; see also Cannon 1981) shows diffuse 
extensions emanating from the elliptical galaxy at position angle $\approx$ 
30$^{\circ}$ almost perpendicular to the major axis of the dusty disk in 
the centre (Fig.~8). The northeastern extension is narrower, longer and 
better-defined than the one in the southwest. It is just west from the 
northern middle radio lobe. The faint extensions most likely consist of stars, 
or possibly of dust reflecting light from stars further in (Cannon 1981; 
Haynes et al. 1983).

\begin{figure}[t]
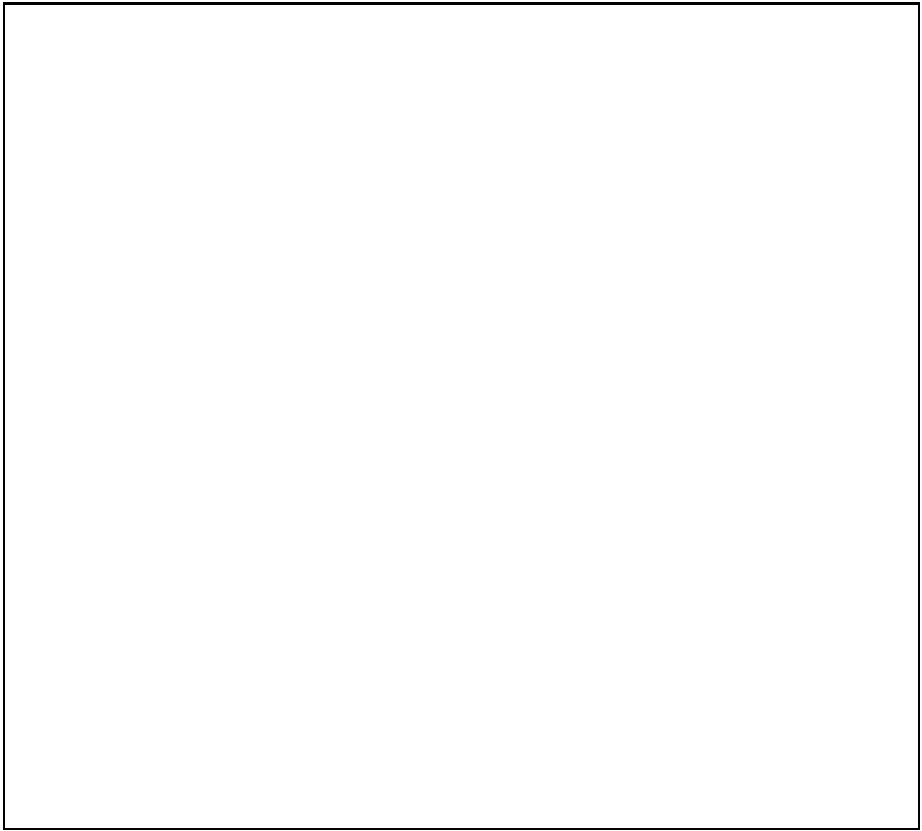
 
\picplace{11.0cm} 
\caption[]{Very deep, amplified positive B-band image of NGC 5128 shows the 
elongated structure of the galaxy, as well as faint emission extending 
roughly along the direction of the inner radio jets. The `traditional' 
negative image of NGC 5128 has been superposed for comparison. The dark band is
clearly inside the galaxy. From Haynes et al. 1983; image courtesy D. 
Malin, Ango-Australian Observatory).
}            
\end{figure} 

The appearance of NGC 5128 departs strikingly from that of a normal elliptical 
galaxy because of its broad and patchy equatorial dark band. Bisecting the 
almost circular bright central part of the galaxy, it is oriented along the 
{\it minor axis} of the elongated shape seen in deep images (Fig.~8). 
Consequently, NGC 5128 is sometimes also classified as a polar ring galaxy 
although it fails to share several of the characteristics common to these (see 
Richter et al. 1994). The dark band is associated with young stellar objects 
(Dufour et al. 1979 -- see also Fig.~6). Modelling has shown that it is in
fact a {\it thin}, strongly warped disk embedded in the host galaxy that 
creates the superficial appearance of a broad band (Sects.~4 and 6). 
Kinematically, the galaxy and the dark band represent different entities. The 
elliptical system and its globular cluster retinue have low rotational 
velocities, whereas the dust disk exhibits much higher rotational velocities 
(Sect.~6.1). 

\subsection{Globular clusters and metallicity}

The galactic foreground confusion caused by the low galactic latitude ($b$ 
= +19$^{\rm o}$) of NGC 5128 for a long time impeded attempts to identify 
its globular clusters, notwithstanding its proximity. However, after the 
first identifications were finally made by Graham $\&$ Phillips (1980) and 
van den Bergh, Hesser $\&$ Harris (1981), the number of confirmed globular 
clusters rose steadily from 20 (Hesser et al. 1984) to 35 (Hesser, Harris 
$\&$ Harris 1986) to 87 (Harris et al. 1992) while Minitti et al. (1996) 
added another 26 globular clusters in the inner 3 kpc of the galaxy. Analyzing 
image counts, Harris et al. (1984a) have estimated a total cluster population 
of 1550$\pm$350. Globular cluster studies have been used primarily to extract 
information on the galaxy's metallicity and dynamics, as well as its distance 
(Sect.~1.2). In terms of both numbers and metallicities, the globular 
cluster system of NGC 5128 appears to be normal for large elliptical galaxies 
(Harris et al., 1984b, 1992) or indeed for spheroidal components of galaxies 
in general (Jablonka et al. 1996).

At least at radii beyond 4$'$, the surface density of globular clusters 
appears to follow the same r$^{1/4}$ law as found by van den Bergh (1976) for
the spheroidal halo light. In addition, the distribution of globular clusters
on the sky hints at a preferential orientation, with a major axis aligned with
the major axis of the outer isophotes of NGC 5128 (Hesser et al. 1984).
Although the first determinations of very high metallicity by Frogel 
(1984) turned out to be overestimates, the clusters in NGC 5128 do seem
to have somewhat greater metallicities ($-0.6 \leq {\rm [Fe/H]} \leq$ +0.1) 
than their Milky Way counterparts which they otherwise resemble 
(Harris et al. 1992; Jablonka et al. 1996; Minniti et 
al. 1996; Alonso $\&$ Minniti 1996). Rather similar metallicities were derived 
for red giant branch stars by Soria et al. (1996). The innermost clusters are 
on average more metal-rich, implying the presence of a metallicity gradient 
$\Delta {\rm [Fe/H]}/\Delta R$ = $-0.08$ dex kpc$^{-1}$; this gradient is 
only apparent, and in fact caused by {\it different concentrations} of 
metal-poor and metal-rich clusters (Alonso $\&$ Minniti 1996, but see 
Jablonka et al. 1996). 

Half a dozen of the innermost clusters have magnitudes and colours (but 
undetermined metallicities) suggesting that they are intermediate-age 
clusters such as found in the Magellanic Clouds (Minniti et al. 1996; Alonso 
$\&$ Minniti 1996). Hui et al. (1995) showed that the metal-poor globular 
cluster ensemble lacks significant rotation whereas the metal-rich ensemble 
rotates more rapidly. A puzzling, but still uncertain result was obtained
by Hesser et al. (1986). They found that the metal-rich innermost cluster 
ensemble has a mean velocity about a 100 $\kms$ (4$\sigma$) higher than the 
systemic velocity of NGC 5128, in contrast to the outermost clusters that 
conform to the systemic velocity. 

The bimodal globular cluster population of NGC 5128 is of great interest 
in view of the proposed merger-nature of the galaxy, particularly as a 
bimodal metallicity distribution with metal-rich clusters more concentrated 
than metal-poor clusters has been identified by Zepf $\&$ Ashman (1993) as 
the natural consequence of galaxy mergers.

\subsection{Late-type objects}

A survey of [OIII] $\lambda$5007 emission from planetary nebulae in NGC 5128 
by Hui et al. (1993a), extending over 20 kpc along the major axis and fully 
covering the central 10 kpc, yielded 785 detections (Hui et al. 1993b). The 
high-luminosity cut-off of the resulting PN luminosity function was used to 
derive the galaxy distance in Sect.~1.2. Hui et al. (1995) measured radial 
velocities for 433 PN's, which were used to study the kinematics and 
dynamics of the galaxy (Sect.~6.2). 

A V-I colour-magnitude diagram of 10\,000 red giant branch (RGB) stars in the 
halo of NGC 5128 was constructed by Soria et al. (1996) from HST/WFPC2 images, 
the first time that individual stars were resolved in a spheroid system 
beyond the Local Group. Analysis of the results yielded the distance and 
metallicity values already mentioned (Sects.~1.2 and 3.2). The I-luminosity 
function of these stars is very similar to that of the RGB stars in the 
M 31 dwarf elliptical companion NGC 185. About 200 stars were found to be 
brighter than the tip of the RGB; most of them are probably upper asymptotic 
giant branch (AGB) stars although confusion with unresolved multiple stars
is still a problem. The luminosity functions of these AGB stars 
suggest the presence of an intermediate-age population of about 5 gigayear, 
but making up at most 10 $\%$ of the total halo stellar population. 

Over the five-year period 1985--1989, Ciardullo et al. (1990) monitored
NGC 5128 for the appearance of novae. They detected 16 novae, twelve of 
which formed a statistically complete and homogeneous sample unaffected by 
the dust lane. Normalized to the near-infrared K-band luminosity, the 
derived nova rate of 4.2 yr$^{-1}$ per 10$^{10}$ $L_{\rm K}$ is virtually 
identical to that derived for the bulge of M 31. They further argued that 
the normalized nova-rate of galaxies is largely independent of their 
luminosity, colour or Hubble type.

The galaxy was the host of the type Ia supernova 1986g (Evans 1986), which 
reached a maximum at $B$ = 12.45$\pm$0.05 (Cristiani et al. 1992). With 
$A_{\rm B}$ = 4.4 mag and the currently best value for the distance modulus, 
this translates into an absolute magnitude $M_{\rm B}$ = $-19.6\pm0.5$. 
However, the reddening of SN 1986g, mostly caused by dust internal to NGC 
5128, is variously given as $E(B-V)$ = 0.9 (Phillips et al. 1987), 1.1 
(Cristiani et at. 1992), 0.6 (Phillips 1993) and 1.6 (Hough et al. 1987). 
The latter also argued from their polarization measurements that near SN 1986g 
the ratio of total to selective extinction R is 2.4, rather than 3.1. The 
resulting $A_{\rm V}$ values thus range from 1.9 to 3.8 mag, so that the 
extinction correction remains a major source of uncertainty. This could be 
resolved by assuming an SNIa mean peak absolute magnitude. Unfortunately, 
estimated values range from $M_{\rm B}$ = $-19.75$ (Cristiani et al. 1992), 
$-19.4$ (Branch et al. 1996) to $-18.2$ (Phillips 1993), the major 
uncertainty this time being galaxy distance moduli. A further discussion of 
the use of SNIa's as standard candles is outside the scope of this review; 
for recent contributions see e.g. Riess, Press $\&$ Kirshner (1996) and 
Branch, Romanishin $\&$ Baron (1996). 

\subsection{Colours and reddening}

The integrated colour of NGC 5128 as given in the RC2 is $(B-V)_{\rm T}$ = 
+0.98. This is identical to the colour van den Bergh (1976) finds at distances 
of 150$''$ to the dark band centerline, where also $(U-B)$ = +0.49. The 
foreground reddening is $E(B-V)$ = 0.11$\pm$0.02 (Newell et al. 1969; Frogel
1984; Harris et al. 1992), so that the intrinsic colours become 
$(B-V)_{\rm o}$ = 0.88 and $(U-B)_{\rm o}$ = 0.41, bluer by 0.08 mag resp. 
0.15 mag than normal for elliptical galaxies (van den Bergh 1976). The 
photometry by van den Bergh shows {\it increasing blueness} approaching the 
dark band; at its edge, 40$''$ from the centerline, the colours abruptly 
spread over a great range +0.4 $\leq (B-V) \leq$ +1.2 and $-0.6 \leq (U-B) 
\leq$ +0.5, whereas in the dark band only {\it red colours} are found 
(+1.0 $\leq (B-V) \leq$ +1.6; +0.3 $\leq (U-B) \leq$ +0.9). 
The excess blue emission has been explained as due to optical synchrotron 
emission from the inner lobes (Dufour et al. 1979), a metal-poor population, 
or a young population of hot, blue stars. The first explanation is unlikely 
(Ebneter $\&$ Balick 1983), as is the second in view of the metallicity 
results given in Sect.~3.2, particularly those by Soria et al. (1996). The 
last explanation is supported by the presence of a substantial number of HII 
regions in the dust band itself (Dufour et al. 1979; Hodge $\&$ Kennicutt 
1983), and even more so by colour images of the galaxy that show substantial
very blue stellar populations at the northwestern and southeastern edges
of the dust band ({\it cf.} Fig.~1). 

A much less sharply delineated and less extreme increase of blue colours 
occurs farther out, in the inner halo, and probably reflects the increasing 
presence of metal-poor objects (van den Bergh 1976).

\section{The dusty disk}

\subsection{Overview}

Dufour et al. (1979) first established that the dark band crossing the
elliptical galaxy is in fact the image of a highly inclined {\it disk 
component} consisting of a metal-rich population of stars, nebulae and dust 
clouds (Fig.~1). Metallicities are close to those in the Solar
Neighbourhood (Dufour et al. 1979; Phillips 1981; Eckart et al. 1990a; 
Viegas $\&$ Prieto 1992). The disk is at position angle 122$^{\rm o}$ $\pm$ 
4$^{\rm o}$ and star formation is rampant. The present burst of star formation 
apparently started 50 million years ago and created at least a hundred HII 
regions embedded in the disk (Dufour et al. 1979; Hodge $\&$ Kennicutt 1983).
Remarkable concentrations of luminous and very blue 
stars can be seen at the northwestern and southeastern edges of the dark band; 
they must represent very large OB associations (Fig.~1). Recent HST
observations allowed Schreier et al. (1996) to also find a large number of 
point-like sources {\it embedded in} the dust band with colours likewise 
suggestive of OB associations (see also Alonso $\&$ Minniti 1996). However, 
since the review by Ebneter $\&$ Balick (1983), little quantitative progress 
has been made on the issue of star formation. This may occur at rates ten 
times higher than in the Milky Way (Telesco 1978) although disk average UV 
energy densities close to Solar Neighbourhood values
(Eckart et al. 1990a) suggest more moderate rates.

As NGC 5128 is no more distant than e.g. M 82, its brightest HII regions and 
supernova remnants should be detectable at centimetre radio wavelengths. Most 
of the published high-resolution radio maps lack the dynamic range to reach 
the low flux-density levels expected for HII regions or SNR's in the disk. 
The 1425 MHz map published by Condon et al. (1996) does, however, show an 
extension with peaks of 125 mJy in an 18$''$ (295 pc) beam coinciding with 
the eastern half of the dark band, while the 43 GHz map by Tateyama $\&$ 
Strauss (1992) likewise seems to show weak emission from the eastern dark 
band. Further high-resolution observations of the disk at centimetre 
wavelengths are desirable as they provide one of the few means 
studying the star formation history of the disk.

As defined by its OB stars and HII regions, the disk extends out to a radius 
of 4 kpc. Molecular line and infrared continuum emission, further
discussed in Sects.~4.2 and 4.3, is concentrated within 40$\%$ of this radius
(Joy et al. 1988; Eckart et al. 1990a; Quillen et al. 1992). HI 
emission, however, extends much farther out, to radii of 7 kpc ({\it cf.} van 
Gorkom et al. 1990; Schiminovich et al. 1994)). The outer parts of the disk, 
as traced by the dark band and the HI emission, show a pronounced warp to 
position angle 90$^{\rm o}$ ({\it cf.} Fig.~7). The disk is in rapid rotation
(Graham 1979; van Gorkom et al. 1990; Quillen et al. 1992). The tilted-ring 
modelling by Nicholson et al. (1992) shows that in spite of appearances, 
the distribution of dust in NGC 5128 is that of a warped {\it thin disk} of 
about 200 pc thickness (Sect.~6.1) along the minor axis. Deep images of NGC 
5128 show the disk to be well inside the elliptical galaxy. Its 
inclination is a function of radius, but remains generally high with respect
to the plane of the sky. The HII 
regions are distributed throughout the warped disk and embedded in diffuse 
ionized gas. Nicholson et al. (1992) showed that their warped disk model 
also quite naturally explains the various CaII and NaI velocity components 
seen in absorption against supernova 1986g by d'Odorico et al. (1989). 
In addition to the seven components associated with NGC 5128, these 
observationsy also showed three components with Galactic foreground gas,
and two intermediate velocity components of unknown origin.

The inner part of NGC 5128 is associated with diffuse X-ray emission 
in the form of ridges along the dark band edges but also in more isotropically 
distributed form (Feigelson et al. 1981; Turner et al. 1997). Although the 
origin of this diffuse emission is not established unequivocally, among the 
most reasonable explanations for its existence are gas ejected from late-type 
stars dynamically heated to the required temperatures, X-ray binaries 
associated with the young stellar population, stellar winds in HII regions 
or combinations thereof (Feigelson, 1981; Turner et al. 1997). In any case, 
NGC 5128 is underluminous in diffuse X-ray emission as compared to other 
early-type galaxies (D\"obereiner et al. 1996).

\subsection{Atomic and molecular gas}

The HI observations by van Gorkom et al. (1990) and Schiminovich et al.
(1994) show the atomic hydrogen to follow the dust lane, including the warp
(Fig.~7). It could, however, not be traced over the central 2.5 kpc because 
of strong absorption against the centre. Van Gorkom et al. (1990) found a 
total HI amount of about 3.3 $\times $10$^{8}$ $\Msun$, but cautioned that 
they might have missed a significant amount because of limited sensitivity;
nor does this estimate include the HI in the shells found by Schiminovich et 
al. 1994 (Fig.~7). Indeed, Richter, Sackett $\&$ Sparke (1994) find within 
the 21$'$ (21 kpc) beam of the Green Bank 140 ft telescope a higher mass of 
8.3$\pm$2.5 $\times$ 10$^{8}$ $\Msun$, still uncertain because of the strong 
central absorption (Sect.~7).

In the central part of the disk, molecular line emission from CO and its 
isotopes is found out to radii of about 2 kpc, but most of it is concentrated 
within a radius of 1 kpc (Phillips et al. 1987; Eckart et al. 1990a; Quillen 
et al. 1992; Rydbeck et al. 1993). Within $R$ = 1 kpc, the area filling factor 
of the disk is of the order of 3--12$\%$, its thickness is less than 35 pc
and the velocity dispersion is about 10 $\kms$ (Quillen et al. 1992). The 
$J$=2--1/$J$=1--0 temperature ratios of about 0.9 for both $^{12}$CO and 
$^{13}$CO as well as the isotopic emission ratios $^{12}$CO/$^{13}$CO = 11 and 
$^{12}$CO/C$^{18}$O = 75 are comparable to those of Milky Way giant molecular 
cloud complexes (Wild, Eckart $\&$ Wiklind 1997). Modelling the CO 
observations as tracer for the much more abundant H$_{2}$ molecule, Eckart et 
al. (1990a) and Wild et al. (1997) estimate molecular hydrogen temperatures 
$T_{\rm k}$ = 10--15 K and densities of a few times 10$^{4}\,cm^{-3}$. 
Emission from other molecular species has also been detected in the disk 
(Whiteoak, Gardner $\&$ H\"oglund 1980; Seaquist $\&$ Bell 1988; d'Odorico 
et al. 1989; Israel 1992; Paglione, Jackson $\&$ Ishizuki 1997).

Total molecular hydrogen masses are probably about 4$\times$ 10$^{8}$ $\Msun$, 
but may be a factor of two higher depending on the CO-to-H$_{2}$ conversion 
factor favoured. The vibrationally excited warm H$_{2}$ ($T_{\rm k}$ $\approx$ 
1000 K) detected by Israel et al. (1990) represents only a minute fraction of 
all molecular hydrogen and is associated with the circumnuclear disk 
(Sect.~5.2). The total gaseous mass of the disk, including helium, is thus of 
the order of 1.3 $\times$ 10$^{9}$ $\Msun$, only about 
2\% of the dynamical mass contained in the elliptical component within the 
radius of the disk ($R$ = 7 kpc). However, because of the pronounced 
concentration of interstellar gas at smaller radii, that fraction increases 
to about 8\% at $R$ = 2 kpc.

\subsection{Dust emission}

At far-infrared wavelengths, the disk of NGC 5128 stands out by its emission 
from warm dust. Dust temperatures are 30--40 K depending on the assumed 
dust emissivity $Q_{100}$ $\propto$ $\lambda ^{-2}$ or $\lambda ^{-1}$ 
respectively (Joy et al. 1988; Marston $\&$ Dickens 1988; Eckart et al. 
1990a). The overall distribution of far-infrared emission is very similar 
to that of the carbon monoxide. The present far-infrared information still 
leaves considerable room for improvement. The KAO scans presented by Joy et 
al. (1988) do not fully sample the galaxy, but show that 10$\%$ of the total 
far-infrared luminosity arises in central source which may be identified with 
the circumnuclear disk (see Sect.~5.2). IRAS survey observations cover the 
whole galaxy, but the resolution is limited. Several of the published fluxes 
refer to poorly calibrated data or underestimate the total flux from the 
extended galaxy. Best fluxes are probably the colour-corrected values $S_{12}$ 
= 26.4 Jy, $S_{25}$ = 25.7 Jy, $S_{60}$ = 236 Jy and $S_{100}$ = 520 Jy given 
by Rice et al. (1988). Use of IRAS non-survey data (DSD maps: Marston $\&$ 
Dickens 1988; CPC-maps: Eckart et al. 1990a; Marston 1992) provided some 
improvement in resolution, but at the cost of photometric accuracy. Because 
of unsolveable calibration problems ({\it cf.} van Driel et al. 1993), the 
image-sharpened CPC maps discussed by Eckart et al. (1990a) and by Marston 
(1992) must be considered as unreliable. The most recent image-sharpened IRAS 
maps (resolution 1--2$'$) incorporating all survey-instrument data are shown 
in Fig.~9; they still lack the resolution to bring forth the full detail of 
the disk.

Good mid-infrared imaging of the dust disk itself is still lacking. The
mean 12$\mu$m surface brightness of the disk as derived from IRAS data 
is about 25 MJy sr$^{-1}$. In addition to the nucleus (Sect.~5.3), Telesco 
(1978) detected 10$\mu$m emission weaker by a factor of about five
from a number of disk HII regions. Clearly, much more remains to be done.

\begin{figure}[t]
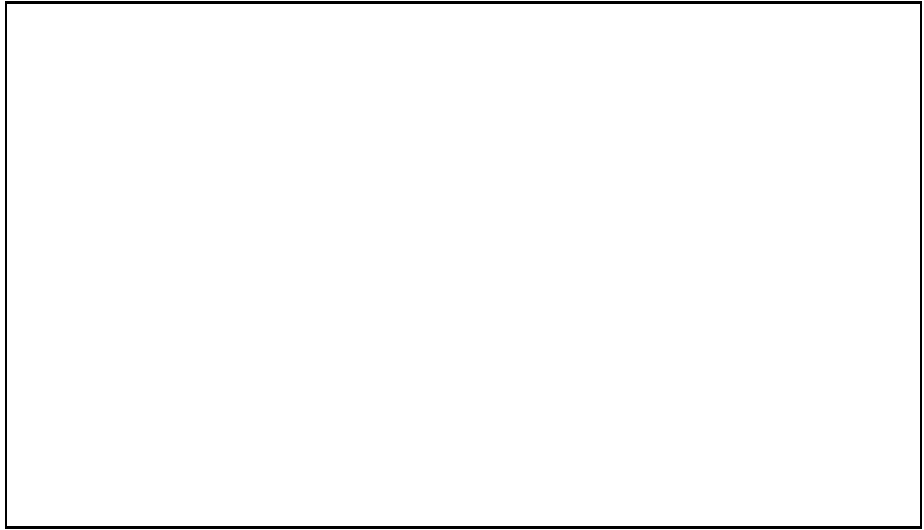
 
\picplace{7.0cm} 
\caption[]{IRAS image-sharpened maps of the NGC 5128 dusty disk at
12$\mu$m (Band 1) and 60$\mu$m (Band 3), showing the warped outer edges.
Courtesy D. Kester, University of Groningen.
}            
\end{figure} 

As much as 50$\%$ of the 100$\mu$m emission may be due to `cirrus' (Marston 
$\&$ Dickens 1988; Eckart et al. 1990a). The various far-infrared data 
indicate the presence of small amounts of dust outside the disk in the main 
elliptical galaxy as well as large amounts in the disk out to a radius of 3 
kpc (Eckart et al. 1990a; Marston 1992). The total dust mass can be estimated 
from IRAS photometry as $M_{\rm d}$ = 1--2 $\times$ 10$^{6}$ $\Msun$ ({\it cf.} 
Hunter et al. 1989) with a luminosity $L_{\rm FIR}$ = 2 $\times$ 10$^{10}$ 
$\Lsun$ (Joy et al. 1988; Eckart et al. 1990a). With considerable uncertainty, 
the {\it total} gas-to-dust ratio within a radius of 2 kpc is 
$M_{\rm gas}/M_{\rm dust}$ = 450. This is an upper limit if significant
amounts of cold dust are present, imperfectly sampled by IRAS. Marston $\&$ 
Dickens (1988) explicitly modelled the IRAS emission in terms of large, cool 
grains and warm, small grains, and arrived at a (distance-corrected) dust 
mass an order of magnitude higher, implying a gas-to-dust ratio of 45, 
which seems rather low. 

\subsection{Polarization and extinction}

At optical and infrared wavelengths, the disk is significantly polarized 
up to 6$\%$ parallel to the dust band (Elvius $\&$ Hall 1964; Hough et al. 
1987; Scarrott et al. 1996). The observations by the latter show higher
levels of polarization at the dust band extremities, with directions 
perpendicular to the dust band. Hough et al. (1987) concluded that the dust 
grains sampled are about 20$\%$ smaller than those in the Milky Way, implying 
an extinction law differing from that in the Solar Neighbourhood, with a 
total to selective extinction ratio of 2.4. The HST R- and I-band imaging 
polarimetry presented by Schreier et al. (1996) shows the polarization to 
reach a peak at a knot close to the nucleus, shining by reflected light. 
Like Hough et al. (1987), they assumed scattering to be negligible elsewhere, 
but that is inconsistent with the apparent importance of scattering in the 
central region and is unlikely in view of the conclusions reached by Packham 
et al. (1996 -- see Section 5.3). The observations by Scarrott et al. (1996) 
can only be 
explained by assuming simultaneous operation of both dichroic extinction 
and scattering, with the latter dominating at the dust band extremities.
This suggests that the very blue colours observed by van den Bergh (1976) 
just at the {\it northern} dark band edge are caused by the light of 
intrinsically blue objects ({\it cf.} Alonso $\&$ Minniti 1996 and references 
therein) enhanced by scattered light and suffering relatively little 
extinction. This is supported by the optical continuum observations of
the central region by Storchi-Bergmann et al. (1997) who find that the
major contribution comes from a metal-rich old bulge, but that there are
also significant contributions from young stars and from scattered light
especially at the dark band edges.

The optical colours suggests significant extinction in the dust band itself 
($E(B-V) \geq$ 0.5 mag -- van den Bergh 1976; Dufour et al. 1979). Indeed, 
the near-infrared estimates by Harding et al. (1981) yield $A_{\rm V}$ = 
3--6 mag, while HST observations indicate V-band extinctions ranging from 
0.5 to 7 mag in the dark band (Schreier et al. 1996) and even reaching values 
in excess of 30 mag ($A_{\rm K} \geq$ 3 mag) just south of the optically 
invisible nucleus (Alonso $\&$ Minniti 1996). The R-band and I-band images 
by Schreier et al. (1996) clearly show how the dust band is seen nearly 
edge-on, but slightly tilted with the near side south of the centre so that 
we are looking from above. The glow of the nuclear region (but not the 
nucleus itself) on the north-side of the dust disk is strikingly apparent 
even through the high extinction it suffers.

\section{The central region}

\subsection{Overview}

The central region of the galaxy, within a few hundred parsec from the
nucleus, is very complex. The nucleus itself, hidden behind thick dust
clouds, is visible at infrared (Sect.~5.3), (sub)millimetre and centimetre 
wavelengths (Sect.~5.4) and again at high energies (Sect.~5.5), but generally 
requires very high resolutions to separate it from its surroundings. 
At centimetre wavelengths 
and at high energies, confusion is caused by the emission from the inner 
jet and even more so from the milliarcsec nuclear jets 
(Meier et al. 1989; Jones et al. 1996; Tingay et al. 1998). 
With increasing observing frequency, the steep nonthermal spectrum of the 
jet features reduces them to insignificance, so that shortwards of 1 cm 
($\geq$ 30 GHz) only the nucleus shines brightly. 

Infrared and millimetre observations have established the presence of a 
compact circumnuclear disk (Israel et al. 1990, 1991; Rydbeck 1993; Hawarden 
et al. 1993; Sect.~5.2).  At wavelengths shorter than 1 mm ($\geq$ 300 GHz), 
thermal emission from this compact circumnuclear disk becomes a new and 
important source of confusion (Cunningham et al. 1984; Hawarden et al. 1993). 
Although at the wavelengths and resolutions at which the nucleus itself can be 
seen, absolute flux-density determinations suffer from various calibration 
problems, it is clear that it is strongly time-variable (Sect.~5.6). 
The highly collimated nuclear radiojets likewise vary their structure and 
intensity with time (Jauncey et al. 1995; Tingay et al. 1998). 

\subsection{The circumnuclear disk}

Analyzing CO measurements and infrared data from the literature, 
Israel et al. (1990; 1991) discovered a circumnuclear disk at the core 
of NGC 5128, with an estimated gas mass 10$^{7} \Msun$. Its outer radius 
is 110--280 pc and it contains a central region of radius 40 pc, devoid of
CO. Such a 100 pc-scale disk appears to be a common feature of active 
galaxies ({\it cf.} Maiolino $\&$ Rieke 1995). 
The CO gas in the disk has excitation temperatures of the order of 25 K and 
is significantly warmer than the CO gas in the dust band ($T_{\rm ex}$ = 
10--15 K (Sect.~4.2). The kinematic signature of this rapidly rotating 
circumnuclear disk is evident in central CO profiles (Israel et al. 1991), as 
well as in major-axis position-velocity diagrams ({\it cf.} Fig.~2 of Quillen et 
al. 1992). The total dynamical mass within the disk area is estimated at 
about 10$^{9}$ $\Msun$, i.e. a hundred times higher than the molecular gas 
mass (Israel et al. 1991; Rydbeck et al. 1993). 
{\it Emission} from other molecular species in the circumnuclear disk,
such as HCO$^{+}$, HCN, HNC and probably also C$_{3}$H$_{2}$, HNCO and 
H$_{2}$CO, has also been detected (Israel 1992).

Although the disk may have been revealed by its near-infrared extinction 
(Meadows $\&$ Allen 1992), it is the observations at 450$\mu$ and 800$\mu$m by 
Hawarden et al. (1993) that provided its first direct image. Its major axis 
is at position angle $140^{\circ}--145^{\circ}$ (Hawarden et al. 1993; 
Rydbeck et al. 1993), quite different from that of the dust band, but 
perpendicular to the position angle of the jets (Sects.~ 2.4 and 5.4). The 
unresolved source seen in limited-resolution far-infrared (Joy et al. 1988)
and mid-infared observations (Fig.~9) is due to thermal emission from
this disk, as the flat-spectrum nonthermal nucleus is expected to contribute 
no more than about 25\%--30\% to the total observed emission at 100$\mu$m.

The disk appears to be larger in the $J$=1--0 transition than in the $J$=2--1 
transition ({\it cf.} Rydbeck et al. 1993). At submillimetre wavelengths it 
is larger
again than in CO or at 100$\mu$m continuum. This suggests a temperature 
gradient across the disk, in addition to the density gradient postulated by 
Israel et al. (1990). Such a temperature gradient is to be expected if the 
excitation of the circumnuclear disk is caused primarily by winds and 
high-energy radiation from the power-law nucleus impinging on the cavity walls.
The very red near-infrared colours, with their implicit suggestion of elevated
dust temperatures in the inner region (Sect.~5.3) in addition to high 
extinction, as well as the compactness of [FeII] and H$_{2}$ emission from 
the centre (Israel et al. 1990; Meadows $\&$ Allen 1992) support this picture. 
Such near-infrared line emission is commonly interpreted as evidence 
either for shocks or for strong X-ray/UV irradiation, also required 
by X-ray observations (Feigelson et al. 1981; Turner et al. 1997; see 
Sect.~5.5). Mouri (1994) has suggested that the H$_{2}$ emission, 
with a temperature of about 1000 K, is excited by UV radiation from the 
nucleus, implying local densities in excess of 10$^{5}\,cm^{3}$ (Sternberg 
$\&$ Dalgarno 1989). This is supported by the detection of X-ray 
line emission and absorption indicating the presence of variably ionized 
absorbing clouds vigorously interacting with the nuclear continuum radiation 
(Morini et al. 1989; Sugizaki et al. 1997; Turner et al. 1997).

\begin{figure}[t]
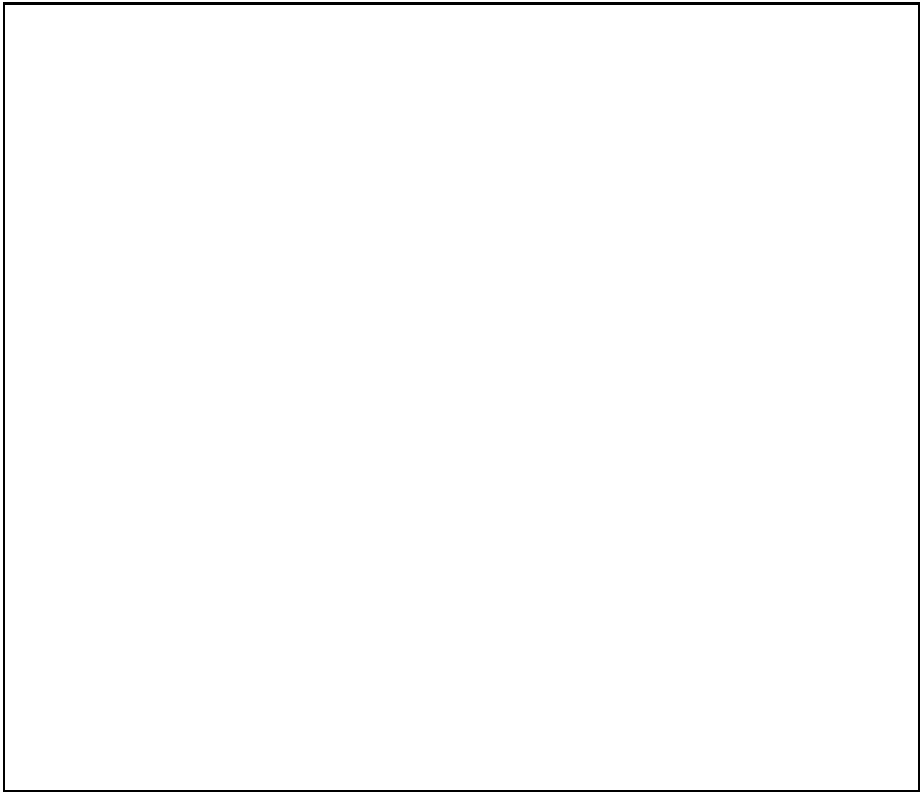
 
\picplace{10.5cm} 
\caption[]{Velocity-position map of the distribution of excited molecular
hydrogen at the inner edge of the circumnuclear disk. The bright central
source indicates rapid rotation close to the nucleus. Courtesy P. van der
Werf, University of Leiden.
}            
\end{figure} 

Unpublished observations by P. van der Werf (private communication) show the 
outer radius of hot, vibrationally excited H$_{2}$ emission to be about 
40 pc where it is rotating with 215 $\kms$ (Fig.~10). This implies
a dynamical mass of 4.4 $\times$ 10$^{8}$ $\Msun$ within 40 pc, or an almost 
twentifold increase of mass per unit volume going from $R$ = 150 pc to $R$ =
40 pc. Assuming the outer limit of the extent of excited H$_{2}$ emission
to mark the cavity 
radius (Koornneef $\&$ Israel 1996), it is doubtful whether the nucleus of
NGC 5128 is powerful enough to provide the required excitation by radiation
alone (Israel et al. 1990). Unless disk material extends much further inwards, 
it cannot control the collimation of the nuclear jets either, 
because this occurs on significantly smaller scales (see Clarke et al. 1992; 
Jones et al. 1996). Whether or not there is another, much smaller 
and thinner accretion disk of the sort proposed by e.g. Krolik $\&$ Lepp 
(1989), it should be emphasized that the agent collimating 
the jets must be closely, if possibly indirectly, connected with the 
circumnuclear disk discussed here in order to explain the excellent alignment 
between its rotation axis and the observed jet flow.

\subsection{Nuclear emission at infrared wavelengths}

Optically, the core of NGC 5128 is completely obscured. Because dust 
extinction in the near-infrared is much less than at optical wavelengths 
(e.g. $A_{\rm K} \approx 0.09 A_{\rm V}$), the core has been frequently 
observed (Becklin et al. 1971; Kunkel $\&$ Bradt 1971; Grasdalen $\&$ Joyce 
1976; L\'epine, Braz $\&$ Epchtein 1984) as well as mapped (Harding, Jones 
$\&$ Rodgers 1981; Quillen, Graham $\&$ Frogel 1993; Adams, Adamson $\&$ 
Giles 1983; Giles 1986; Joy et al. 1991; Meadows $\&$ Allen 1992; Turner et 
al. 1992; Golombek, quoted in Schreier et al. 1996) in the 1--5$\mu$m
wavelength range. Near-infrared images longwards of 2$\mu$m, usually show 
a compact source close to the central position, but it is questionable 
whether this represents direct emission from the nucleus (see e.g. 
Antonucci $\&$ Barvainas 1990). 

The relatively large-scale K-band maps by Quillen et al. (1993) and Adams 
et al. (1983) show a central 2.2$\mu$m peak embedded in an extended emission 
component in addition to several foreground stars. The extended emission
is mostly due to the innermost stars of the elliptical galaxy, although there 
is also some emission from stars and HII regions in an inner dust lane of 
outer radius of 80$''$ and projected thickness of 15$''$ (Harding et al. 1981; 
Quillen et al. 1993; Adams et al. 1983; Meadows $\&$ Allen 1992).
These K-band images show the `blossoming' of emission northeast of a dust 
absorption band, visible even at this wavelength, that extends across the 
nucleus. Below 2$\mu$m, the central feature is cometary in shape, points to 
the southwest in the same position angle as the X-ray/radio jet (Sect.~2.4) 
and is less compact than at longer wavelengths. The northeastern part of this 
feature corresponds to the UBVRI `hot spot' discovered by Kunkel $\&$ Bradt 
(who prematurely identified it with the nucleus). Joy et al. (1991)
speculated that this cometary J/H-band `blue' source is the near-infrared 
counterpart of the X-ray/radio jet discussed in Sect.~2.4. However, its 
`blue' infrared colours and high degree of polarization are consistent with 
reflected light from a stellar or active nucleus in a clear line of sight
({\it cf.} Turner et al. 1992; Packham et al. 1996; Schreier et al. 1996), 
whereas the opening angle of the `blue' component is much wider 
than that of the radio/X-ray jets.

The position of peak emission moves to the southwest with increasing 
wavelength ({\it cf.} Schreier et al. 1996). The accuracy of positional 
measurements
is insufficient to determine whether or not the K-band peak coincides with the 
nucleus (for the best position determinations see Giles 1986), but data 
tabulated by Schreier et al. (1996) suggest it is still somewhat off.
Near-infrared polarimetry of the visually obscured centre of NGC 5128 
has shown that, with increasing wavelength as well as with decreasing 
aperture, the position angle of the polarized vectors steadily increases from 
$PA$ $\approx$ 120$^{\rm o}$ to $PA$ = 145$^{\rm o}$ (Bailey et al. 1986;
Packham et al. (1996), i.e. from parallel to the dust band to parallel to 
the circumnuclear disk and perpendicular to the radio jet. At the `nuclear
source', the intrinsic polarization is 17$\%$. Schreier et al. (1996) further 
showed that this polarization peak coincides with a compact knot close to both 
the extinction and K-band emission peaks. The knot appears to be a heavily 
obscured interstellar cloud, scattering the optical and near-infrared 
radiation of the invisible nearby nucleus toward our line of sight. 
The absence of significant polarization at {\it millimetre} wavelengths led 
Packham et al. (1996) to reject Bailey et al.'s (1986) notion of an 
intrinsically highly polarized nucleus, emitting near-infrared synchrotron 
radiation and instead led them to conclude that the nucleus itself is obscured 
even at near-infrared wavelengths with $A_{\rm V}$ = 16 mag and that the 
`nuclear' polarization is produced by scattering and not by synchrotron 
emission or dichroic extinction. 

Thus, the true nucleus is only fully revealed at wavelengths longer than 
2.2$\mu$m, and the compact K-band source is mostly due to reflected light. 
Indeed, the extinction of the {\it nucleus itself} is variously estimated to 
be $15 \leq A_{\rm V} \leq 60$, centering on $A_{\rm V}$ = 25--30 (see 
L\'epine et al. 1984; Giles 1986; Turner et al. 1992; Meadows $\&$ Allen 1992, 
and references therein). As this is much higher than the extinction caused
by the dust band (Sect.~4.4), the nucleus must be obscured primarily by
the circumnuclear disk. Apart from the extinction problem, it is extremely 
difficult to determine reliable flux-densities, colours or even the very 
reality of weak unresolved K-band sources in galaxy centres, because of the 
practical difficulties in separating such sources from the central cusp of the 
surrounding galaxy (Simpson 1994; see also Fig.~8 by Turner et al. 1992). 

The nucleus of NGC 5128 does appear as an infrared point source at 3.3
$\mu$m (Turner et al. 1992) and at 10$\mu$m (size $\leq$ 1$''$ i.e. $\leq$ 
16 pc; flux 1.4 Jy -- P. van der Werf, private 
communication); it also is variable at least at 3$\mu$m (L\'epine et al. 
1984; Turner et al. 1992; see Sect.~5.6). Observations at longer infrared
wavelengths so far have lacked the resolution to accurately separate the 
nucleus from the circumnuclear disk emission.

Infrared spectra centered on the nucleus show emission lines due to
molecular hydrogen (H$_{2}$), Brackett-$\gamma$ and [FeII] (Israel et
al. 1990; Meadows $\&$ Allen 1992), but lack the noticeable CO-band
absorption (characteristic for late-type stellar populations) which
is seen off the nucleus where, in turn, the emission lines are absent
(Meadows $\&$ Allen, 1992). The near-infrared colours of the central
area, in contrast to those of the `blue feature' mentioned above, are
extremely red, going from $J-H$ = 1.5, $H-K$ = 0.9 in a 2.5$''$ aperture
(Turner et al. 1992) to $J-H$ = 1.35, $H-K$ = 1.75 in a 1$''$ aperture
(Giles 1986). The very red colours are confined to a region within
12$''$ from the nucleus (Giles 1986). They are far too red be caused
by extinction. A rather similar situation has been found for NGC 3079 by 
Israel et al. 1998 who interpret such colours as due to a combination of 
extinction and emission from hot (1000 K) dust grains. Reddened synchrotron 
emission, also considered by Turner et al. (1992) appears to be ruled out 
by the work of Packham et al. (1996) discussed above.

\subsection{The radio nucleus and nuclear jet}

At the core of Centaurus A lies a compact radio nucleus for which 
(Kellerman, Zensus $\&$ Cohen (1997) have measured a size of 0.5$\pm$0.1 
milli-arcsec corresponding to linear dimensions of only 0.008 pc or 1700 AU). 
As there may be unresolved fine-scale structure even at this resolution,
actual nuclear dimensions could be even smaller, the {\it lower limit} 
being about 10$^{16}$ cm or 700 AU (Grindlay 1975; Mushotzky et al. 
1978; Jourdain et al. 1993). 

The nuclear source has a strongly inverted spectrum ($\alpha \approx$ 4; Jones 
et al 1996). It is all but invisible below 5 GHz, but shows up in VLBI maps 
at 8.4 GHz with significantly time-variable flux-densities ranging over 
an order of magnitude (Preston et al. 1983; Meier et al. 1989; Jauncey et al. 
1995; Jones et al. 1996; Tingay et al. 1998). Below about 20--30 GHz the radio 
emission from the nucleus probably suffers from both synchrotron 
self-absorption and free-free absorption in a circumnuclear ionized gas 
(Jones et al. 1996). Above the turnover frequency, the spectrum appears to 
remain flat (flux-density about 8--10 Jy up to at least 1000 GHz ({\it cf. }
Kellerman et al. 1997). 

Very high resolution (VLBI) radio measurements are needed to separate the 
extremely compact nucleus from its surroundings. Over some 65 milli-arcsec 
(projected linear distance 1 pc) a bright linear jet can be traced emanating 
from the nucleus at a position angle of 51$^{\circ}$ and with a width of a few 
milli-arcsec (Preston et al. 1983; Jones et al. 1996). The VLBI maps by Jones 
et al. (1996) and Tingay et al. (1998) show a weak counterjet to 
the bright northeastern nuclear jet. The jet components have very similar 
spectral indices $\alpha = -0.77$ and, like the nucleus, are
variable in intensity (Meier et al. 1989). Monitoring the 
jet over several years, Tingay et al. (1998) found significant 
structural (c.q. intensity) changes of the knots in the jet (Fig.~11), 
implying internal evolution as well as a subluminal projected motion $v$ 
$\approx$ 0.1c of the knots. 

\begin{figure}[t]
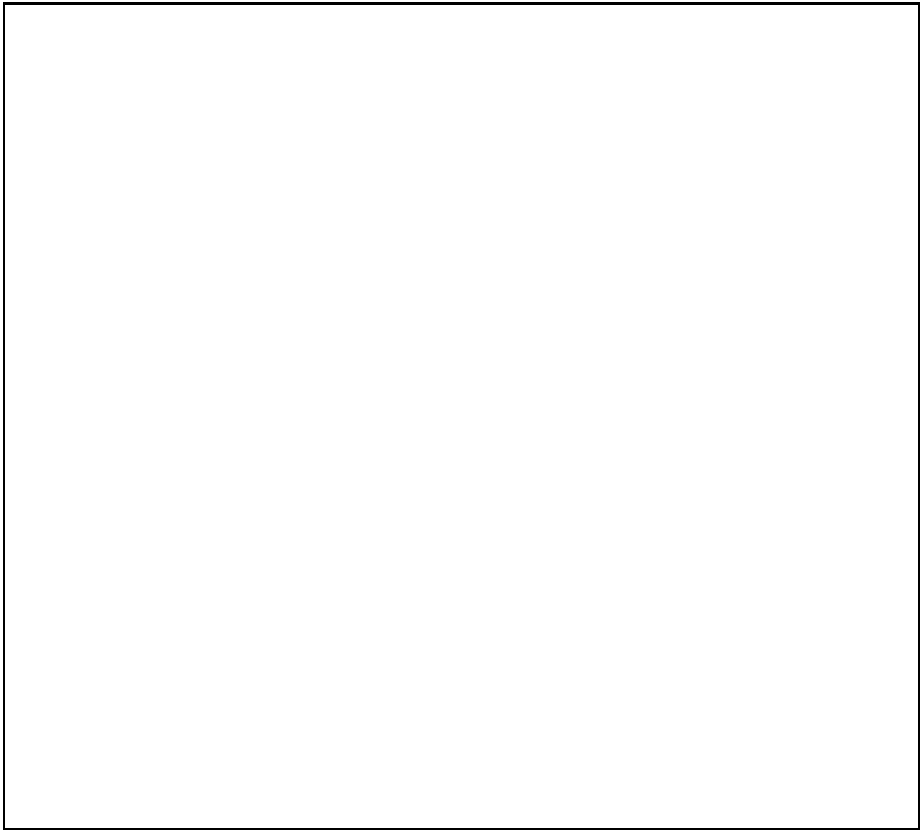
 
\picplace{11.0cm} 
\caption[]{VLBI maps of the core and nuclear jet at 8.4 GHz over the period
1991--1996 identifying major components, and showing jet motion. From
Tingay et al. (1998).
}            
\end{figure} 

This {\it observed motion} may in fact represent slow patterns superposed
on a significantly faster relativistic jet flow (Tingay et al. 1998).
The large difference in jet and counterjet brightnesses can be explained 
by relativistic doppler beaming, enhancing the radiation of the jet 
approaching us. From the observed brightness ratio of 4--8, it appears that
the northeastern jet is approaching and the southwestern jet receding at an 
angle of 50$^{\circ}$--80$^{\circ}$ to the line of sight at moderately 
relativistic speeds $v$ $\geq$ 0.45c (Jones et al. 1996; Tingay et al. 1998;
see also Bao $\&$ Wiita 1997). The two brightest knots C1 and C2 separated 
recently from the nucleus, in 1983 and in 1989 respectively (Tingay et al. 
1998).

There is still a significant gap in radio coverage of the nuclear region 
between aperture synthesis maps (5 GHz {\it resolution} 0.3$''$: Clarke et al. 
1992) and VLBI maps (8.4 GHz {\it coverage} $\approx$ 0.1$''$: Jauncey et al. 
1995; Tingay et al. 1998) so that a full description of the nuclear jets is 
still lacking. Since the nuclear jet velocities are much higher than those 
estimated for the inner jet (Sect.~2.4), significant deceleration must take 
place just at the missing size range so that any possibility of filling the 
observational gap should be pursued.

\subsection{X- and $\gamma$-ray emission from the nuclear region}

The hard X-ray/soft $\gamma$-ray emission is thought to arise from Compton 
up-scattering of lower energy photons by relativistic electrons; the
photons may be supplied by the relativistic electrons themselves via
the synchrotron self-Compton process, or by accretion heating. Major 
outbursts, correlated at X-ray and radio wavelengths, would then represent 
injection of fresh relativistic 
electrons, or re-acceleration of Compton-cooled electrons. They should 
decline on timescales determined by synchrotron loss rates, i.e., a few 
years. Faster variations probably reflect source expansions (Grindlay 1975; 
Mushotzky et al. 1978; Baity et al., 1981; Skibo et al. 1994; {\it cf.} Bond 
et al. 1996). Unlike the nucleus, at least the outer jets radiate primarily at
soft X-ray energies (Feigelson et al. 1981; Turner et al. 1997)

The best available X-ray resolution, 6$''$ provided by
the ROSAT HRI, is insufficient to separate the nucleus and the nuclear jets,
and even the innermost part of the inner jet. The energy distribution of the 
nucleus has been characterized by power-law spectra (Ubertini et al. 1993
Jourdain et al. 1993; Miyazaki et al. 1996; Sugizaki et al. 1997; Turner et 
al. 1997) with exponential cutoffs at 200--300 keV during periods of increased 
activity and 700 keV during quiescent periods (Kinzer et al. 1995). Centaurus 
A is difficult to detect above energies of 1 MeV (see Kinzer et al. 1995 and 
references therein) where the BL-Lac object MS1312-4122, 2$^{\circ}$ west of 
Centaurus A is also a potentially serious source of confusion (Fichtel et al. 
1994; Kinzer et al. 1995; Thompson et al. 1995). 
In particular, detections in the 
1--20 MeV range by von Ballmoos et al. (1987) were not confirmed by O'Neill 
et al. (1989), although the latter observed the nucleus in a more intense 
state. Likewise, an early detection at very high energies (Grindlay et al. 
1975) could not be repeated (Carrami\~nana et al. 1990; {\it cf.} Allen et al. 
1993).
However, observations with COMPTEL have now provided good detections 
in the 1--30 MeV range (Steinle et al. 1998), which together with a 200 MeV 
EGRET detection (Thompson et al. 1995) allow the spectrum to be fitted with 
sets of broken power laws, with spectral indices steepening from 
$\alpha = -1.7$ to $-2.0$ to $-2.6$ for the low ($L_{\gamma}$ = $3 
\times 10^{42}$ erg s$^{-1}$) intensity state and from $\alpha = -1.7$ 
to $-2.3$ to $-3.3$ for the intermediate ($L_{\gamma}$ = $5 \times 10^{42}$ 
erg s$^{-1}$) intensity state.

Turner et al. (1997) concluded that nuclear emission dominates the X-ray flux
above 2 keV and estimated that 40\% of the source suffers extinction by a 
column $\nh$ = 4$\times$10$^{23} \cm2$, 59\% by a column $\nh$ 
= 1$\times$10$^{23} \cm2$ and the remaining 1\% by a column two orders of 
magnitude less. It is tempting to ascribe the first fraction to the nucleus 
itself, and the second to the nuclear jets, but there is no real evidence 
supporting this speculation. It is clear, in any case, that both components
are obscured by the circumnuclear disk discussed in Sect.~5.2 whose
estimated parameters fit the X-ray observations rather well. X-ray K-edge
absorption and fluorescent emission from Fe at 6.4 keV (Mushotzky et al. 1978; 
Wang et al. 1986 Morini et al. 1989), but also Mg, Si and S (Sugizaki et al. 
1997) can be explained by irradiation and reprocessing of clouds in the 
circumnuclear disk by the nuclear X-ray source ({\it cf.} Morini et al. 1989; 
Turner et al. 1997) and thus provide independent evidence for the fierce
interaction of the nucleus with the inner parts of the circumnuclear disk.

\subsection{Variability of the nucleus}

Both the nucleus and the jet structures within at least 0.1$'$ are 
variable. As they do not seem to vary in tandem (Tingay et al. 1998) and 
have different spectra, accurate determinations of the spectral 
index and its time variation are extremely difficult to make especially 
with relatively large single-dish beams. 
At 1.4 GHz, considerable variability is seen on timescales of a hundred 
days or longer (Romero, Benaglia $\&$ Combi 1997). As the amplitude of 
variation is large, and the nucleus is strongly (self)absorbed at this 
frequency (Sect.~5.4), it cannot reflect activity in the nucleus itself.
Instead, shocks interacting with 
density inhomogeneities in the inner jet appear responsible. Well-established
variability at higher frequencies is harder to interpret. Large-beam 
monitoring by Botti $\&$ Abraham (1993) has yielded large flux (22 GHz: 16--32 
Jy; 43 GHz: 6--20 Jy) and spectral index variations ($\Delta \alpha$ 
= 2), while Kellerman et al. (1997) noted a 60\% drop in 43 GHz intensity 
over only 3 months. Qualitatively, this may reflect expansion of the 
nuclear source and jets, initially opaque and then becoming (partially) 
optically thin. However, the large beams used include significant non-nuclear 
emission from the nuclear and inner jets, confusing the issue.

Nevertheless, the radio variability appears to be correlated with that at 
hard X-ray wavelengths (Botti $\&$ Abraham 1993; Jourdain et al. 1993; Turner 
et al. 1997) which must be associated with the nucleus.
The 3--12 keV light curve presented by Turner et al. (1997) for a 26-year 
period (based largely on the Vela 5B light-curve from Terrell 1986, and
sparsely sampled after 1980) shows major outbursts in 1972--1976 and in 
1979 (see also Feigelson et al. 1981; Baity et al. 1981; Gehrels et al. 1984; 
Bond et al. 1996) when the observed X-ray fluxes increased by an order of 
magnitude. These outbursts were separated by a low-intensity state with 
fluctuations by a factor of two. At millimetre wavelengths, likewise sampling 
nuclear behaviour, the 1972--1976 maximum is evident in the 90 GHz
measurements by Kellermann (1974) and Beall et al. (1978). The 90--800
GHz measurements by Cunningham et al. (1984; see also Hwarden et al. 1993) 
show that the outburst starting in 1979 was still going on in 1981,
which is also confirmed at 100 keV (Gehrels et al. 1984; Jourdain et al. 1993).
A poorly covered outburst occurred in 1985 (Jourdain et al. 1993; Bond
et al. 1996; Turner et al. 1997); this outburst may be associated with the
expulsion of radio knot C1 (Tingay et al. 1998). The near-infrared 3.5$\mu$m 
variability noted by L\'epine et al. (1984) and Turner et al. (1992) follows a
pattern similar to that of the radio and X-ray emission (low in in 1971 and 
1987, high in 1975 and 1981) suggesting that the emission at this wavelength 
arises from a similar or identical mechanism.

The observed 0.1--2 keV luminosity of the nuclear region increased by a 
factor of two between 1990 and mid-1993. A probably related rise was seen in 
the 43 GHz radio flux occurring at the end of 1992 lasting into early 1993
(Abraham 1996). However, no flux increase was seen at hard X-ray energies 
(Turner et al. 1997; Bond et al. 1996) whereas 100/230 GHz 
monitoring by Tornikoski et al. (1996) reveals only modest ($\leq 50\%$) 
flux increases over relatively short times in 1992/1993 and 1994/1995.
As both hard X-rays and millimetre wave frequencies sample the nucleus 
rather than the jets, this suggests that the soft X-ray/43 GHz increase 
is related to the jet expansion observed by Tingay et al. (1998) and not 
to nuclear activity. 

The hard X-ray flares of July/October 1991 (Jourdain et al. 1993; Bond et al.
1996) were unfortunately not covered by the millimetre monitoring
programme. The October 1991 flare is evident, however, in $\gamma$-ray 
observations by Kinzer et al. (1995); see also $\gamma$-ray fluxes listed 
by Steinle et al. (1998) for the period 1991--1995, and their graph of the 
hard X-ray variability over the same period showing significant activity in 
mid-1994.
Minor X-ray flares in the low-intensity state have doubling times of about two 
days and last for one or two weeks; they appear to occur several times a year 
(Terrell 1986; Beall et al. 1987; Jourdain et al. 1993; Turner et al. 
1997) and involve energy transfers of order 10$^{42}$ erg s$^{-1}$ 
(Terrell 1986; Bond et al. 1996). The nucleus seems to have been in the 
quiescent, low-intensity state for most of the time since 1989, as also 
indicated by the 8.4 GHz VLBI core monitoring by Tingay et al. (1998).
In addition to these variations, there is at least at X-ray and soft
$\gamma$-ray wavelengths a considerable small-amplitude (10--20$\%$) 
variation at timescales of minutes and hours (Morini et al. 1989; Kinzer et 
al. 1995; Miyazaki et al. 1995; Turner et al. 1997, but see also Jourdain 
et al. 1993). Very fast flux variations reported by Wang et al. (1986) have 
not been confirmed and are now generally doubted.

\section{Kinematics and dynamics}

\subsection{Kinematical studies}

Over the last two decades, several major kinematical studies have been
carried out of both the ellipsoidal and the disk component of NC 5128.
The kinematics of the ellipsoidal component have been traced by studies
of the motions of stars (Wilkinson et al. 1986), globular clusters
(Hesser, Harris $\&$ Hesser 1986; Harris, Harris $\&$ Hesser 1988) and 
planetary nebulae (Hui et al. 1985).

Stellar motions were fully mapped by Wilkinson et al. (1986) over 100$''$ 
and supplemented by major axis data out to 400$''$ (6.6 kpc). The ellipsoidal 
stellar component is rotating rather slowly at velocities not surpassing 40 
$\kms$ in the line of sight around an axis at position angle 135$^{\circ}$, 
i.e. rotating only roughly perpendicular to the dust lane. The velocity 
dispersions range from 95 to 150 $\kms$. The planetary nebula system, studied 
by Hui et al. (1995) out to 20$'$, has a similar mean velocity dispersion of 
110 $\kms$. Major axis rotation increases to 100 $\kms$ at 7 kpc and after 
that remains constant to at least 22 kpc. Minor axis rotation is less than 50 
$\kms$ out to the last observed radius of 10 kpc. The 
projected rotation axis is at position angle 165$^{\circ}$, i.e. displaced 
from the photometric minor axis (and disk orientation) by about 40$^{\circ}$;
the line of zero rotation is not orthogonal to the line of maximum 
rotation. Such observations suggest that NGC 5128 has a triaxial potential. 
The kinematics of the globular cluster system again appear to be similar to 
those of the spheroidal stellar system (Hesser et al. 1986; Harris et al. 
1988). In particular, the metal-rich globular clusters rotate very much like 
the planetary nebulae, whereas the metal-poor clusters appear to lack 
significant rotation (Hui et al. 1995). 

The rich gas content of the dust lane (Sects.~ 4.1 and 4.2) has allowed 
detailed kinematical studies of the disk component to be performed in lines 
of ionized hydrogen (Bland, Taylor $\&$ Atherton 1987), neutral atomic 
hydrogen (van Gorkom et al. 1990; Schiminovich et al. 1994) and CO (Eckart 
et al. 1990a; Quillen et al. 1992; Wild et al. 1997). 
Optical and radio observations (Graham 1979; van Gorkom et al. 1990; Quillen 
et al. 1992) show the disk to be in rapid rotation with a rotation gradient 
of about 150 $\kms$ kpc$^{-1}$ close to the nucleus, in sharp contrast to 
the rather modest rotation of the elliptical component. Graham (1979) 
estimated the plane of rotation to be tilted by 73$^{\rm o}$ 
with respect to the plane of the sky. 
Bland et al. (1987) mapped the ionized hydrogen gas at arcsec resolution over 
7$' \times$5$'$, with a moderate velocity resolution of 36 $\kms$ similar to
that obtained by Wilkinson et al. (1986); the neutral hydrogen observations
have comparable velocity resolutions. Although the CO observations have 
high velocity resolutions of about 1 $\kms$, that advantage 
is largely undone by their spatial resolution of 20$''$--40$''$ causing 
significant beam-smearing which can only be partially undone by modelling. 

The limited resolutions of the HI and CO data allow only relatively simple
modelling of circular orbits in warped disks approximated by sets of 
variously tilted rings. Virtually all attempts to dynamically model the
observed kinematics of the disk therefore rely largely or completely on
the optical observations by Bland et al. (1987). Assuming circular motion 
consistent with an r$^{1/4}$ law mass distribution, Nicholson, Bland-Hawthorn 
$\&$ Taylor (1992) confirmed Graham's (1979) conclusion that the gas and dust 
are in a highly inclined rotating disk, with an amplitude $V_{\rm rot}$ = 250 
$\kms$ at $R$ = 2.0 kpc and a rotation gradient similar to the one derived by 
Graham (1979), eastern edge approaching. Projected onto the sky, the HII 
regions are confined to an envelope similar tothe form of a hysteresis loop, 
characteristic of a warped disk
rather than a ring. They convincingly showed that the kinematical 
data can be represented by a class of models involving a thin warped 
disk geometry with a spatially averaged optical depth sufficiently low to 
allow emission from all positions of the disk to be seen. Double-peaked 
velocity profiles mark folds in the warped disk seen in projection. 
Quillen et al. (1992; 1993) confirmed that such an interpretation also 
applies to the dusty molecular disk. 

\subsection{Dynamical models}

Kinematical information is restricted to the distribution on the sky of
radial velocities only. The large number of free parameters in combination 
with observational limitations such as finite resolution and extinction, 
renders the outcome of dynamical modelling  very sensitive to initial
assumptions ({\it cf.} Kormendy $\&$ Djorgovski 1989). It is therefore not
surprising that the general agreement on kinematical (tilted ring) models 
describing the motion of the disk component is not matched by a 
similar convergence on a unique dynamical model. For a review of the methods 
and problems of determining structure and dynamics of elliptical galaxies in 
general the reader is referred to De Zeeuw $\&$ Franx (1991).

Wilkinson et al. (1986) found a ratio of maximum rotational velocity to peak 
nuclear velocity dispersion characteristic of a rotationally supported oblate 
system seen almost edge-on (Davies et al. 1983). Assuming the radio jet to 
mark the long axis, an effectively stationary rotating nearly oblate triaxial 
model with axis ratios 1:0.98:0.55 provided their best fit; our line of sight 
is then 30$^{\circ}$--40$^{\circ}$ from the plane defined by the two long 
axes. Dropping the requirement of identical jet direction and stellar symmetry 
axis, they also found a satisfactory fit to an effectively stationary prolate 
spheroidal model. Following earlier work by Tubbs (1980), a prolate model 
with clouds moving in circular orbits in a warped disk was 
favoured by Quillen et al. (1992) on the basis of their CO observations and 
the ionized gas kinematics determined by Bland et al. (1987). Quillen et al. 
(1993) presented a revised version of this model to reproduce the morphology 
of their near-infrared maps.

Assuming the planetary nebulae to form a relaxed system (implying a
triaxial potential), a stationary galaxy figure, and the gas disk to be
in the preferred plane defined by the long axis, Hui et al. (1995) derived
a model with axial ratios of 1:0.92:0.79 where the short axis lies in the 
plane of the sky, and the intermediate axis is along the line of sight.
Mathieu, Dejonghe $\&$ Hui (1996) used the planetary nebula system to
construct triaxial models, assuming a spherical potential obtained by
inverting the major axis photometry. Failing to obtain satisfactory fits
with constant M/L ratios, they concluded to a significant presence of
dark matter, causing a significant increase of the mass-to-light ratio 
with radius. They consider the kinematics and photometry of NGC 5128
to be well described by two kinematically distinct subsystems supporting
the merger hypothesis. On the whole, there is substantial agreement that the
inner parts of NGC 5128 have a roughly constant mass-to-light ratio
$M/L_{\rm B}$ $\approx$ 4 increasing with radius to the extent that
the {\it global} ratio of NGC 5128 is $M/L_{\rm B}$ = 10 (Hesser et al.
1984; van Gorkom et al. 1990; Hui et al. 1995; Mathieu et al. 1996).
The total mass of the galaxy is $M$ = 4$\pm$1 $\times$ 10$^{11}$ M$\sun$,
with typically half of it due to dark matter (Mathieu et al. 1996).
The galaxy mass is thus between 2$\%$ and 10$\%$ of the NGC 5128/NGC 5236
group mass.

Interpreting the dust band as a precessing warped structure,
Nicholson et al. (1992) ruled out prolate models and concluded that its 
geometry is consistent with a near-polar gas disk in an oblate, almost 
spherical triaxial potential with its short axis likewise in the plane of the 
sky. In their models, the dust band is separated into an inner detached disk 
($r < $1.7 kpc) and an outer extended annulus (1.7 kpc $ < r <$6.8 kpc). 
Sparke (1996) showed that, at least qualitatively, such models can 
explain not only the the observed disk features but also the 
broken HI ring observed by Schiminovich et al. (1994). An important factor is 
the variation in oblateness of the galaxy, which is taken to be nearly 
spherical at small radii and more flattened farther out. Sparke's (1986) 
model resembles that of Wilkinson et al. (1986) but is in contradiction to 
that of Hui et al. (1995). The discrepancy may be resolved if, in fact, the 
planetary nebulae do not represent a relaxed and well-mixed system (Sparke 
1996).
 
Of particular interest is the shell system referred to in Sect.~4.1 and 
shown in Figs.~6 and 7. Malin et al. (1983) interpreted the shells as the 
signature of the collision between a dynamically cold system and a rigid 
potential well (Quinn, 1984; Hernquist $\&$ Quinn 1988, 1989) under the 
condition that the infalling system had a mass significantly less than that 
of the elliptical galaxy. The shells arise from phase-wrapping of the disk 
after it has been disrupted by the tidal field of the massive elliptical
galaxy. In fact, this work, as well as that by Tubbs (1980) 
served to underpin the galactic-encounter interpretation of NGC 5128 first 
suggested by Baade $\&$ Minkowski (1954) and revived by Graham (1979). 
Malin et al. (1983) suggested that a late-type galaxy of mass a few times 
10$^{10}$ $\Msun$, similar to e.g. M 33, merged with NGC 5128 a few hundred 
million years ago. Specific, more detailed merger scenarios were presented 
by Tubbs (1980) and Quillen et al. (1993).

\subsection{Evolutionary timescale}

If the gas dust disk would define a preferred plane in a triaxial system, 
its rotation should be retrograde with respect to the tumbling motion of
the stellar ellipsoid (van Albada, Kotanyi $\&$ Schwarzschild 1982). 
However, Wilkinson et al. (1986) found that the disk has in fact prograde 
rotation, and suggested the warped disk to be due to incomplete settling of 
material into a symmetry plane of the potential. This is consistent with the
conclusion by van Dokkum $\&$ Franx (1995) that dust disks in early-type
galaxies, with a radius exceeding 250 pc, are generally not settled. The disk
would thus be a transient phenomenon, unless it were stabilized by 
self-gravity (Nicholson et al. 1992; see also Sparke 1996).

A stable disk is required if the merger and dust band formation occurred 
some 10$^{9}$ years ago (Graham 1979; Nicholson et al. 1992). Quillen et 
al.'s (1992; 1993) `short' timescales of 1--2 $\times$ 10$^{8}$ years were
criticized by Sparke (1996) who estimated three quarters of a 
gigayear as the time elapsed since capture of a companion galaxy from a 
polar orbit. The presence of an intermediate age stellar population in the 
halo of NGC 5128 prompted Soria et al. (1996) to suggest an even longer 
timescale of several gigayear, while intermediate scales of 2--8 $\times$ 
10$^{8}$ years (and incomplete settling) have been suggested by Tubbs (1980) 
and Malin et al. (1983). All these timescales are, however, much longer than 
the age of the current burst of star formation in the disk which is typically 
a few times 10$^{7}$ years (van den Bergh 1976; Dufour et al. 1979)

\section{The nuclear absorption spectrum}

HI line absorption against the centre of Cen A over the velocity range
of 500--600 $\kms$ was first recognized in single-dish observations by 
Roberts (1970) and Whiteoak $\&$ Gardner (1971; 1976b). It was studied 
at high spatial (2.6$'' \times$11$''$) and moderate velocity
(6 $\kms$) resolutions by Van der Hulst, Golisch $\&$ Haschick (1983).
Subsequently, centimetre wavelength absorption by 
various molecular species (OH, H$_{2}$CO, C$_{3}$H$_{2}$, NH$_{3}$) was
observed by Gardner $\&$ Whiteoak 1976a, b; 1979; Seaquist $\&$ Bell 1986; 
1988; 1990). The HI observations by van der Hulst et al. (1983) show the 
inner jets (20$''$ and 30$''$ northeast of the nucleus) as well as the
nuclear source in absorption. The inner jets are also seen in H$_{2}$CO 
absorption (Seaquist $\&$ Bell 1990), while H$_{2}$CO absorption is likewise
found in the direction of the dust band, 4$'$ southeast from the nucleus 
(Gardner $\&$ Whiteoak 1976b).

The flat and strong nuclear continuum spectrum at millimetre wavelengths 
(Sect.~5.4) provides the opportunity to extend absorption
studies to wavelengths where molecular line
transitions are most abundant. High velocity resolution ($\Delta V \leq$ 2.6 
$\kms$) absorption spectra of the lower transitions of $^{12}$CO and 
$^{13}$CO were obtained by Israel 
et al. (1990; 1991) and Eckart et al. (1990b). Various other molecular
species, such as HCO$^{+}$, H$^{13}$CO$^{+}$, HCN, HNC, CS, C$_{2}$H, 
CN, C$_{3}$H$_{2}$ and H$_{2}$CO have likewise been detected in
absorption (Eckart et al. 1990b; Israel et al. 1991; Wiklind $\&$ Combes 
1997; Israel et al. unpublished). Centaurus A is unique in this sense:
as of yet, no other active galactic nucleus is known to exhibit absorption
by a molecular disk (Drinkwater, Combes $\&$ Wiklind 1996).

Below 10 GHz, emission from the inner and nuclear jet dominates the continuum, 
whereas above 30 GHz the continuum emission arises mostly or wholly from the 
nucleus. Thus, centimetre-wavelength single-dish measurements sample 
sightlines as far as 500 pc away from the nucleus while aperture synthesis 
observations, after elimination of the inner jet contribution, 
preferentially sample sightlines missing the nucleus itself by about 
1.5 parsec, i.e. 200 times the diameter
of the nuclear source. Only at (sub)millimetre wavelengths does the 
absorption sample a narrow line of sight (0.5 milliarcsec) directly to the
nucleus. 

A typical molecular line spectrum of the Centaurus A nucleus consists of a 
relatively strong emission line arising from the molecular disk 
(Sect.~4) as well as weaker and broader emission representing the
circumnuclear disk (Sect.~4.2) superposed on the nuclear continuum ({\it cf.}
Fig.~1 in Israel, 1992). As absorption occurs over a significant fraction 
of the total emission line spectrum, the intrinsic shape of the latter is 
hard to determine but of crucial importance in determining the pure 
absorption spectrum.

\begin{figure}[t]
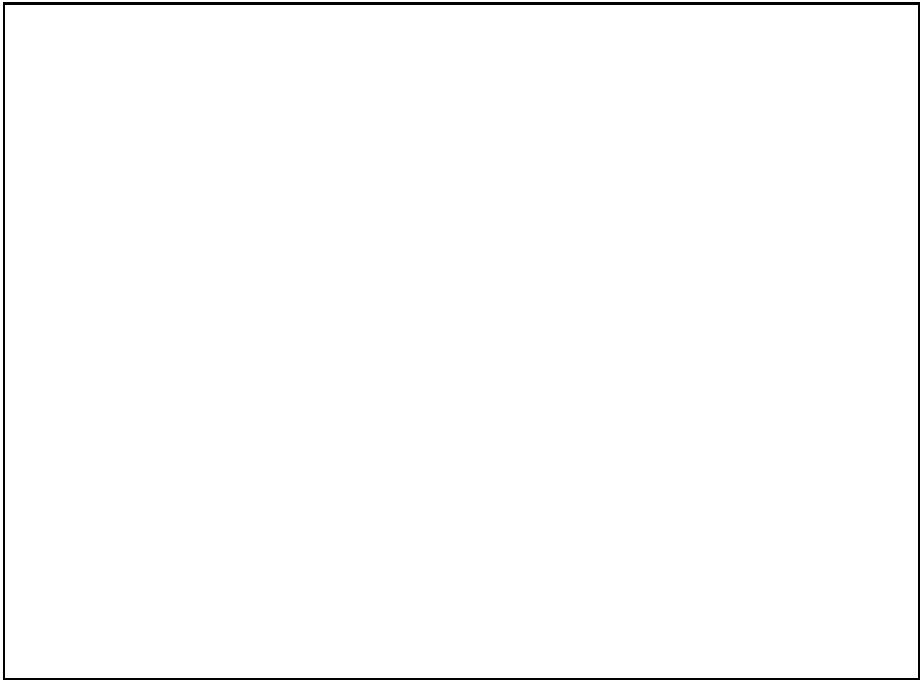
 
\picplace{9.0cm} 
\caption[]{Nuclear absorption spectrum in HCO$^{+}$. Left: low velocity
resolution (4.7 $\kms$) spectrum showing continuum, molecular line emission
from circumnuclear disk and dark band, and molecular line absorption
covering large velocity range. Right: high velocity resolution (0.3 $\kms$)
spectrum showing several narrow absorption lines and `forest' of redshifted 
lines. Horizontal scale is $V_{\rm LRS}$, vertical scale is antenna
temperature (corresponding to 0.75 $\times$ main-beam brightness temperature).
From Israel et al. 1991, and unpublished data.
}            
\end{figure} 

The absorption spectra show numerous peaks at various velocities (Fig.~12),
relatively strong between $540 \kms \leq V_{\rm Hel} <\leq 555 \kms$, i.e.
within 10 $\kms$ from the systemic velocity. It is 
not clear which of these lines, if any, marks the systemic velocity. 
There are at least five (blended) 
components in this velocity range but only two of these have discernible 
HI counterparts in VLA spectra 
(van der Hulst et al. 1983; J.M. van der Hulst, private communication).
As these two HI features are also seen against the inner jets 
at 30$''$ (500 pc) above the circumnuclear disk, they originate most likely 
in the more distant molecular disk associated with the dust band structure
({\it cf.} van der Hulst et al. 1983). This is consistent with 
the apparent low excitation of the corresponding molecular absorption lines, 
which 
suggest the material to be located at considerable distance to the nucleus 
(Gardner $\&$ Whiteoak 1979; Seaquist $\&$ Bell 1990; Eckart et al. 1990b).
The velocity range is consistent with the velocity dispersion in the 
molecular disk (Quillen et al. 1992).

Of particular interest is a second absorption line system, extending from
$V_{\rm Hel}$ = 560 $\kms$ to 650 $\kms$. It is seen only against the
nuclear source, and not against the inner jets (disregarding the marginal 
H$_{2}$CO feature seen by Seaquist $\&$ Bell 1990). This high-velocity,
redshifted absorption system takes the shape of continous molecular 
absorption with a few individual peaks of relatively low optical depth 
(Israel et al. 1991; Wiklind $\&$ Combes 1997). The redshifted
absorption system is also prominent in HI (Gardner $\&$ Whiteoak 1976b;
van der Hulst et al. 1983). With respect to e.g. HCO$^{+}$, the HI
absorption is relatively strong at 560 $\kms \leq V_{\rm Hel} \leq$ 600
$\kms$, and weak at higher velocities. This redshifted system is
generally interpreted as due to infalling clouds (Gardner $\&$ Whiteoak 
1976b; van der Hulst 1983; Seaquist $\&$ Bell 1990; Israel et al. 1991). 
It implies accretion rates sufficiently high to explain the overall radio
luminosity of Centaurus A (van Gorkom et al. 1989).
As the absorption optical depth ratios of the various species appear
to be a function of (infall) velocity, this opens the interesting
possibility of modelling the processing of material falling into
the nucleus of Centaurus A. As the optical depths and consequently
signal-to-noise ratios are low, this is not an easy task.

At velocities $500 \kms \leq V_{\rm Hel} \leq 540 \kms$ there is a hint of a 
very low optical depth blueshifted HCO$^{+}$ absorption wing ({\it cf.} Israel 1992; 
Wiklind $\&$ Combes 1997). Its nature is not clear, and it is not seen in HI
(J.M. van der Hulst, private communication).
High resolution OH maser absorption spectra against the nuclear source
in the four transitions at 18 cm wavelength show
exactly conjugate behaviour at all velocities in the satellite lines: the
sum of the two transitions is zero (van Langevelde et al. 1995). 
This effect is caused by the competition of the two transitions for the same
pumping photons and allows a direct determination of the OH column density
$N(OH) \approx$ 6 $\times$ 10$^{15} \cm2$.

\section{Concluding summary}

NGC 5128 is a massive elliptical galaxy at the heart of a moderately rich 
group of galaxies (Sect.~1.4); basic data are summarized in Table 3.
Although most of its properties are fairly normal for a luminous triaxial 
elliptical
galaxy, it is remarkable in two aspects: it hosts a very large radio source
(Sect.~2) and its inner parts harbour a relatively massive disk of dust, gas 
and young stars (Sect.~4). Both have been proposed as the consequence of past 
merger activity. The location of the galaxy amidst several dwarf galaxies 
lends plausibility to such a suggestion. Indeed, although no {\it direct} 
evidence of a merger has been found, the appearance of the galaxy and in 
particular the properties of its embedded disk, such as gas mass, kinematics, 
warp and polar orientation along the photometric minor axis ({\it cf.} 
Bertola et 
al. 1988) as well as the bimodality of its globular cluster system (Zepf $\&$ 
Ashman 1993; Perelmuter 1995), the existence of luminous optical and and HI 
shells as well as the outcome of various dynamical models and scenarios 
(Sect.~6.2) all provide strong indirect evidence that at least one major 
merger event occurred some 10$^{8}$--10$^{9}$ years ago. 
The gas mass of the dusty disk, a few times 10$^{9}$ $\Msun$, point at 
capture of a fairly sized late-type spiral galaxy rather than a small 
irregular. The shell structures in particular suggest that NGC 5128 
experienced more than just one merger (Weil $\&$ Hernquist 1996).  

The radio source Centaurus A, associated with NGC 5128, is a very near 
example of a large class of radio
galaxies of moderate luminosity known as FR-I galaxies (Fanaroff $\&$ 
Riley 1974). Radio sources of this class are generally presumed to
have moderately active nuclei with relativistic outflows on a subparsec
scale not aligned with our line of sight. The observations of
Centaurus A at radio and X-ray/$\gamma$ ray wavelengths are 
consistent with this interpretation.  
NGC 5128 contains a very compact nucleus of size 1200$\pm$500 A.U. from 
which subluminal relativistic jets emanate (sect.~5.4) that become 
subrelativistic within 1.5 pc. The jets appear to propagate at a large 
angle to our line of sight. The nucleus itself is strongly obscured by a 
small (radius $\approx$ 150 pc) circumnuclear disk (Sect.~5.2) and is quite 
variable at radio and X-ray wavelengths (Sect.~5.6). The polarization 
of the central region, the ionization of the optical filaments 
and the apparent similarity of the high-energy spectrum (but not the 
luminosity) of Centaurus A (in particular at $\gamma$-ray energies) to 
that of blazars and quasars such as 3C273, have been used to argue that 
the galaxy indeed harbours a misaligned BL Lac/blazar nucleus (Bailey et 
al. 1986; Morganti et al. 1991; Dermer $\&$ Schlickeiser 1993; Kinzer et al. 
1995; see also Steinle et al. 1998, and references therein). The substantially 
lower luminosity of Centaurus A is then explained by our viewing the galaxy 
from the side, and not down the jet axis. However, some caution to this 
conclusion has been expressed by Antonucci $\&$ Barvainis (1990) and 
Kellerman et al. (1997).

\begin{table}[htb]
\centering
\caption[]{NGC 5128 Basic Data }
\begin{tabular}{lrll} 
\hline\noalign{\smallskip}
				&  Value	& Units & Reference \\
\noalign{\smallskip}
\hline\noalign{\smallskip}
$\alpha$(1950)$_{\rm o}$ 	& 13:22:31.6$\pm$0.2 & & Giles 1986 \\
$\delta$(1950)$_{\rm o}$ 	& -42:45:30.3$\pm$0.4 & & Giles 1986 \\
Galactic Longitude $l$		& 309.5		& degrees	& \\
Galactic Latitude $b$		& +19.4		& degrees	& \\
Systemic Velocity $V_{\rm Hel}$ & 543$\pm$2	& $\kms$	& Table 1 \\
Galaxy Size D$_{25}$		& 18$\times$14  & arcmin	& RC2 \\
Radio Source Size		& 8$\times$4	& degrees	& Section 2.1 \\
Distance			& 3.4$\pm$0.15  & Mpc		& Section 1.2 \\
Apparent Magnitude $B$		& 7.96		& mag		& RC2 \\
Colour $(B-V)_{\rm T}$		& 0.98		& mag		& RC2 \\
Foreground Reddening $E(B-V)$   & 0.11		& mag		& Section 3.4 \\
Total Galaxy Mass		& 4$\pm$1 $\times$ 10$^{11}$ & $M_{\sun}$ & Mathieu et al. 1996 \\
Total HI Mass			& 8.3$\pm$2.5$\times$ 10$^{8}$ & $M_{\sun}$ & Section 4.2 \\
Gas Mass Dusty Disk		& 1.3$\pm$0.4$\times$ 10$^{9}$ & $M_{\sun}$ & Section 4.2 \\
Gas Mass Circumnuclear Disk	& 1.1$\pm$0.3$\times$ 10$^{7}$ & $M_{\sun}$ & Section 5.2 \\
\noalign{\smallskip}
Linear Sizes: & & & \\
Outer Radio Lobe		& 250		& kpc		& Section 2.1 \\
Middle Radio Lobe		& 30		& kpc		& Section 2.2 \\
Inner Radio Lobe		& 5		& kpc		& Section 2.3 \\
Inner Radio Jet			& 1.35		& kpc		& Section 2.3 \\
Relativistic Nuclear Jet	& 1.65          & pc		& Jones et al. 1996 \\
Radio Core			& 0.008		& pc		& Kellerman et al. 1997 \\
Radius Dusty Disk		& 7		& kpc		& Section 4.1 \\
Radius Circumnuclear Disk	& 150(-40, +130) & pc		& Section 5.2 \\
\noalign{\smallskip}
\hline
\end{tabular}
\end{table}

Nuclear activity must have been going on for a considerable amount of
time, given the size of the outer radio lobes. The bulk speeds of 5000 $\kms$
estimated for the inner jets (Sect.~2.4) and the outer radius of 250 kpc
of the giant lobes of radio emission suggest {\it a lower limit} of 50
million years. As the inner jets appear to dissolve into plumes (`inner
lobes') at about 5 kpc from the nucleus (Sect.~2.3), and as the position 
angle of the outer radio features is much different from that of the inner 
features, it is reasonable to conclude that significant deceleration occurs 
over most of the radio source, leading to a substantially higher age. 
Indeed, the age of the inner lobes alone was already estimated at
6 $\times$ 10$^{8}$ years (Slee et al. 1983), although this may be 
too high. 
The jets appear to lose much of their energy within a few parsec from the 
nucleus, presumably by interaction with ambient material. The peculiar radio
brightness evolution of component C1 in the nuclear jet may provide
a clue to this process (Tingay et al. 1998) underscoring the need for
further VLBI monitoring of Centaurus A, as well as the desirability of
filling the resolution gap in the 0.1--0.3$''$ range. The inner jets 
dissolve in the more extended inner lobe plumes, which exhibit a profound 
clockwise bending (decreasing position angle). Again, ambient material and 
its movement in the galaxy, may explain the observed morphology (see e.g. 
Sparke 1982; Gopal-Krishna $\&$ Saripalli 1984; Heckman et al. 1985), but
hard evidence is lacking. Moreover, this is unlikely to also 
explain the similarly profound clockwise bending of the giant outer lobes, 
well outside the optical galaxy. Noting a continuous decrease of position
angle (i.e. clockwise bending) of various features at increasing
distance to the nucleus, Haynes et al. (1983) have proposed that
the central collimating source precesses at a rate of the order of
10$^{-5}$ degrees per year. The discovery of a circumnuclear disk
perpendicular to the nuclear and inner jet, yet inclined to the
minor axis of the elliptical galaxy, supports the idea of precession.
If the rate of procession is correctly estimated, the structure of the
radio source should exhibit the effect of several precessional periods.
It would be interesting to see whether
the run of position angles with radius can indeed be modelled by such
a precession of the collimating agent. Alternatively, a combination of
precession and ambient gas dynamics may be required, while the structure 
of the outer lobes, in addition, may be influenced by tumbling and orbital 
motion of the galaxy as a whole (Burns et al. 1983).

The putative age of the merger (Sect.~6.3) suggests a link to the origin of 
the radio source. Although the presence of an active nuclear source 
predating a merger cannot be excluded, it is tempting to associate its origin 
with the accumulation of matter in the centre caused by transfer of angular
momentum through viscous damping after such an event.
An intriguing indication that the origin of the radio source is 
connected to merger activity is provided by a morphological argument.
The bended appearance of the giant radio lobes (Fig.~4) is very similar
to that of the tilted rings forming the dust band ({\it cf.} Fig.~11b in
Nicholson et al. 1992) rotated by 90$^{\circ}$. For instance, the position
angle of 0$^{\circ}$ characterizing the outer radio contours corresponds
to the position angle of 90$^{\circ}$ of the outer rings. Because the
dynamical time scale of the outer rings is much longer than that of the
more strongly tilted inner rings, their {\it present} position angle should 
more closely resemble the {\it original} orientation of the inner disk
structure at the time that the matter now forming the outer lobes
was ejected. As the dust disk originated in a merger event, and the
morphology of the radio lobes appear to follow its subsequent
evolution, it seems likely that the activity creating these lobes
is also a consequence of the merger event.

If the nuclear source is in fact a black hole, 
its estimated bolometric luminosity (half of it at high energies) of
about 10$^{43}$ erg s$^{-1}$ implies a {\it lower limit} to the
black hole mass of about 5$\times$10$^{4} \Msun$ whereas the total 
luminosity of the radio
source suggests a mass $\geq 10^{7} \Msun$ (Terrell 1986 and references
therein; Kinzer et al. 1995). As the dynamical mass within 40 pc is about 
4$\times$10$^{8}$ $\Msun$, and there is no obvious sign of Keplerian 
rotation (see Fig.~10), its {\it upper limit} 
must be a few times 10$^{7}$ $\Msun$. Although the high obscuration of
the centre of NGC 5128 and the lack of H$_{2}$O masers (Braatz et al. 1996) 
precludes, at the moment, a more accurate mass determination, the actual 
mass of the putative black hole nevertheless is fairly well-constrained
and is comparable
to that of the circumnuclear disk (Sect.~5.2). This mass is
not very high and infalling molecular clouds, especially dense cores, 
may penetrate deeply before being tidally disrupted. The variability 
of the nucleus may represent the accretion of individual stellar or cloud 
remnants onto the black hole triggering renewed jet activity (Sect.~5.4
through 5.6) and fueling the radio source. Details of these processes 
are not clear yet, but careful and 
frequent monitoring of Centaurus A at radio, X-ray and $\gamma$-ray 
wavelengths may provide important information.
For instance, how does the nucleus drive the nuclear jets, and how are 
the relativistic nuclear jets transformed into the nonrelativistic inner 
jets? The circumnuclear disk (Sect.~5.2) does not seem capable of 
controlling the collimation of the nuclear jets, but its 
orientation exactly perpendicular to these jets, suggests that it is 
somehow connected with the collimating agent. Comparison of Centaurus A 
features with very-high resolution observations (HST, VLBA) of other 
active elliptical galaxies suffering less nuclear extinction, such as the 
ten times more distant NGC 4261 (e.g. Jones $\&$ Wehrle 1997, and 
references therein) may prove particularly fruitful. 

\medskip

\begin{acknowledgement}
It is a pleasure to thank David Malin, Do Kester, Norbert Junkes, 
Thijs van der Hulst, St\'ephanie C\^ot\'e, Steven Tingay, Jack
Burns and Paul van der Werf for kindly supplying the illustrations 
in this review. I also would like to thank Paul van der Werf, 
Hans Bloemen, George Miley and in particular Tim de Zeeuw 
for critical comments on an earlier version of this work.
The burden of literature searches was greatly relieved by the use of the
NASA Astrophysics Data System (ADS) Astronomy Abstract Service
\end{acknowledgement}

\end{document}